# On-chip transfer of ultrashort graphene plasmon wavepackets using terahertz electronics

Katsumasa Yoshioka[1,*], Guillaume Bernard[1], Taro Wakamura[1], Masayuki Hashisaka[1,2], Ken-ichi Sasaki[1], Satoshi Sasaki[1], Kenji Watanabe[3], Takashi Taniguchi[4], and Norio Kumada[1]

[1]NTT Basic Research Laboratories, NTT Corporation, 3-1 Morinosato-Wakamiya, Atsugi, 243-0198, Japan

[2]Institute for Solid State Physics, University of Tokyo, 5-1-5 Kashiwa-no-ha, Kashiwa 277-8581, Japan

[3]Research Center for Electronic and Optical Materials, National Institute for Materials Science, 1-1 Namiki, Tsukuba 305-0044, Japan

[4]Research Center for Materials Nanoarchitectonics, National Institute for Materials Science, 1-1 Namiki, Tsukuba 305-0044, Japan

*e-mail: katsumasa.yoshioka@ntt.com

## Abstract:

**Ultrashort polariton wavepackets, such as terahertz graphene plasmon polaritons, could be used for fast information processing on integrated circuits. However, conventional optical techniques have struggled to integrate the components for controlling polariton signals and have low conversion efficiency. Here, we show that graphene plasmon wavepackets can be generated, manipulated and read-out on-chip using terahertz electronics. Electrical pulses injected into a graphene microribbon via an ohmic contact can be efficiently converted into a plasmon wavepacket with a pulse duration as short as 1.2 ps and three-dimensional spatial confinement of $2.1 \times 10^{-18}$ m$^3$. The conversion efficiency between the electrical pulses and plasmon wavepackets can also reach 35% due to the absence of a momentum mismatch. The transport properties of graphene plasmons are studied by changing the dielectric environments, providing a basis for designing graphene plasmonic circuits.**

**Main text:**

Graphene plasmon polaritons in the terahertz (THz) and mid-infrared frequency range[1–5] could be used as information carriers in next-generation devices. Parallel to developments in semiconductor plasmonics[6], advances in graphene plasmon polaritons have led to strong light confinement capabilities, high controllability through carrier density modulation, and low loss characteristics[1–5]. Various graphene plasmonic circuits have been proposed for both classical[7–10] and quantum[11–13] information transfer and processing; these circuits feature active functionalities through electrostatic gating that are unattainable with metallic plasmonic circuits[14,15]. The fundamental properties of graphene plasmons have been extensively investigated using optical techniques such as far-field spectroscopy[16–24] and near-field scanning optical microscopy[25–33]. However, these methods measure the absorption of localized plasmon resonance[16–24] or standing waves of plasmons[25–33] and are not directly applicable to operate plasmonic circuits.

To realize the full functionality of graphene plasmonic circuits, methodologies capable of efficiently generating dynamically controllable plasmon signals, manipulating their propagation, and remotely measuring their amplitude and phase on-chip are needed. One of the largest obstacles to achieving this is the momentum mismatch between the incident optical excitation and graphene plasmons. While subwavelength structures, such as antennas and gratings, can overcome this mismatch[16–33], the conversion efficiency between the optical excitation and the graphene plasmon wavepacket is low, making the excitation and detection of propagating plasmons challenging. Furthermore, the subwavelength structures restrict the device geometry.

One strategy to circumvent the momentum mismatch is to inject spatiotemporally localized electrical pulses, which can create a transient local charge density imbalance in the system and subsequently launch plasmons[34]. While high-frequency gigahertz (GHz) electronics have been used to make time-resolved transport measurements of graphene plasmons in bulk[35] and quantum Hall edge channels[36–38], conventional electronics have bandwidth limitations that preclude access to the desired THz range. Therefore, it is necessary to develop electronics and device architectures capable of handling THz plasmons in order to realize graphene plasmonic circuits for fast information transfer and processing.

In this Article, we report efficient electrical generation, manipulation, and detection of propagating ultrashort graphene plasmon wavepackets. This is achieved using on-chip THz spectroscopy[39–44] and integrated circuits consisting of a graphene-embedded coplanar waveguide combined with top-gate structures for controlling the carrier density of graphene. By using electrical excitation via an ohmic contact, we can address the momentum mismatch issue intrinsic to optical excitation, achieving a conversion efficiency between excitation pulse and plasmon of around 35%. The broadband high-frequency circuit response is engineered to enable dynamic tracking on-chip of plasmon wavepackets as short as 1.2 ps. The wavepackets are short enough to allow us to directly investigate crucial metrics such as propagation velocity, spatial confinement, propagation length, and conversion efficiency in graphene microribbons in different dielectric environments.

## Terahertz electronic setup

Our device, illustrated in Fig. 1a, integrates a high-quality single-layer graphene transistor encapsulated with hexagonal boron nitride (hBN) into the center of a coplanar waveguide (CPW). This hBN encapsulation substantially enhances the mobility of the graphene, thereby reducing plasmon dissipation[30]. The transmission region of the CPW comprises a 10-μm-wide center conductor flanked by two ground planes, each separated from the center conductor by a 7-μm gap. Both ends of the graphene form ohmic contacts with the center conductor through titanium/gold (Ti/Au) edge contacts[45]. The graphene was etched into a micro ribbon with a length of over 20 μm and a width of 8 μm. We fabricated two devices with different top-gate structures: one had a ZnO gate and the other had a standard metal (Ti/Au) gate. The ZnO top gate is transparent in the GHz and THz range[23,43]. This enables us to investigate plasmons free from the gate screening (unscreened plasmons) while tuning the carrier density. In the metal gate device, on the other hand, the screening effect gives rise to acoustic plasmons[20–22,28,29,32] with the electric field confined between the gate electrode and graphene. Understanding the transport properties of plasmon wavepackets in graphene microribbons under different screening conditions is critical for constructing plasmonic circuits with tunable functionalities. Detailed descriptions of these structures are presented in the following sections (Fig. 2 for unscreened and Fig. 3 for acoustic plasmon).

We conducted pump-probe experiments using a pulsed femtosecond laser with a 280-fs pulse duration and a wavelength of 1035 nm[46] to avoid photo-induced doping in graphene[43]. The beam was divided into pump and probe beams with a controlled time delay to excite low-temperature-grown gallium arsenide (LT-GaAs) photoconductive (PC) switches placed on either side of the graphene: one PC switch is used to inject a THz electrical pulse into the graphene from one ohmic contact and the other is used to read out the time-domain waveform after it passes through the graphene. Figure 1b shows the waveform of the THz electrical pulse measured using a CPW without graphene. The pulse duration is 1.2 ps at full width at half maximum (FWHM) with the frequency component ranging from 0 to 2 THz, as shown in the inset. This waveform corresponds to the THz electrical pulse used to electrically excite ultrashort graphene plasmon wavepackets throughout the study. All measurements were performed at 4 K in a vacuum.

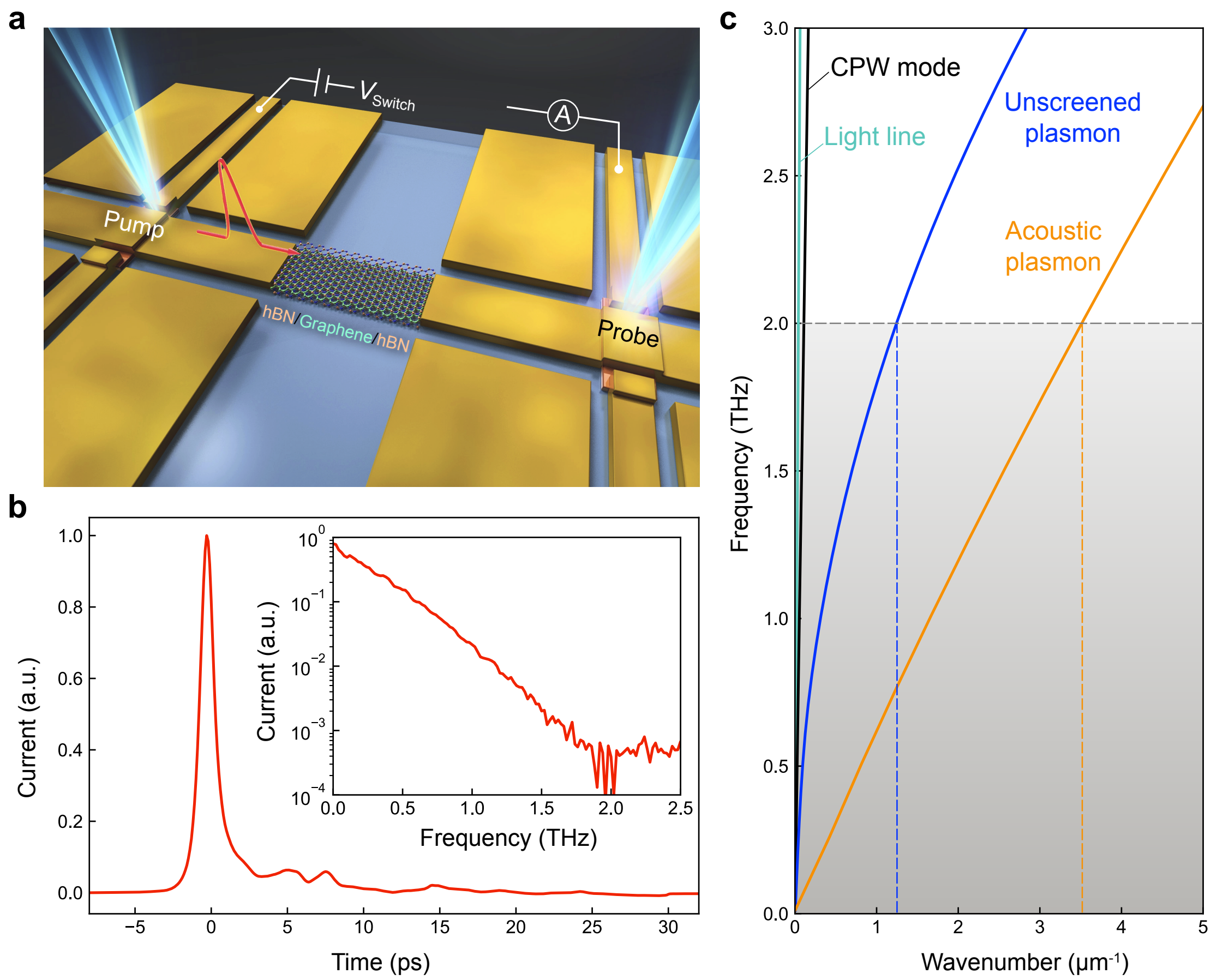

**Figure 1 | Terahertz electronics setup. a**, Schematic device structure. A graphene transistor is integrated into a coplanar waveguide (CPW). Detail gate structures are presented in Figs. 2a, b and Figs. 3a, b. Both ends of the graphene form ohmic contacts with the center conductor of the CPW, facilitating injection and extraction of THz electrical current. The current is generated and measured in the time domain through LT-GaAs PC switches by using a pump-probe method. **b**, Temporal profile of THz electrical current in a CPW without graphene for a travel length of 485 μm. This ultrashort electrical pulse is used to electrically excite graphene plasmons throughout the study. Measurements were performed at 4 K. The inset shows the normalized Fourier-transformed spectrum. **c**, Dispersion curves of graphene plasmon. The gray shaded area highlights the range of frequency components of the input electrical pulse with varying spectral weights. Darker areas denote heavier spectral weights. The dispersion curve of graphene plasmons determines the wavenumber of the generated plasmon wavepacket, with the blue and orange dashed lines marking the largest wavenumbers. CPW mode corresponds to the even mode excited in the CPW (see Supplementary Section III for details).

Figure 1c illustrates our plasmon excitation approach, characterized by spatiotemporally localized charge injection. In this scheme, both the frequency component of the electrical pulse and the dispersion relation of the graphene plasmons jointly determine the wavenumber of the generated plasmon wavepacket. This methodology ensures the consistent excitation of a broadband wavepacket within our on-chip circuitry. The propagated waveform is then accurately captured by a detector PC switch.

## Experimental results

### Unscreened graphene plasmons in a ZnO gate device

Let us begin by investigating the transport properties of unscreened graphene plasmon (Fig. 2c) wavepackets on the ZnO gate device schematically shown in Figs. 2a and b. The graphene in this case was 8-μm wide and 23-μm long. Figure 2d shows the time-domain waveform of the THz electrical pulse as a function of the ZnO gate voltage ($V_{ZnO}$: -3 to 3 V, carrier density $n$: -6.2 to 7.9 × $10^{11}$ $cm^{-2}$). The time origin ($t$ = 0) corresponds to the time at which the THz electrical pulse was applied on the ohmic contact, and the peak position corresponds to the time of flight of the wavepacket through the 23-μm-long graphene micro ribbon (see Supplementary Section I-a for details). The crosstalk can be removed by subtracting the signal at the CNP ($V_{CNP}$ = -0.35 V), where plasmons cannot propagate, from the signals at finite carrier densities (see methods for details). For both electron and hole doping, an ultrashort plasmon wavepacket with a peak position around 1-2 ps appears; its amplitude decreases with decreasing carrier density due to increased plasmon propagation loss and impedance mismatch at higher graphene resistivities. The asymmetry of the amplitude in the n-doped and p-doped regions is due to contact doping[41]; graphene is n-doped near the contacts, which results in diminished plasmon amplitude for p-doped regions due to higher resistance caused by two p-n junctions formed near the contacts. Note that the small signals around 5 and 9 ps originate from multiple reflections: the first between the generation PC switch and graphene over 480 μm, and the second between the detection PC switch and graphene over 960 μm, all at a propagation velocity of 1.2 × $10^8$ m/s in the CPW. Now let us examine the propagation characteristics of the plasmon wavepackets by comparing the waveforms normalized by the peak amplitude in the inset of Fig. 2d. With increasing carrier density, the peak position shifts to an earlier time, indicating increase in plasmon velocity. The velocity calculated by dividing the graphene length (23 μm) by the peak position is plotted in Fig. 2e as a function of the carrier density. The velocity increases with doping from 1.5 to 1.8 × $10^7$ m/s. Alongside this velocity change, the pulse duration decreases, reaching 2.0 ps (FWHM) at $V_{\mathrm{ZnO}}$ = 3.0 V. Despite this shortening, the pulse is still broader than the input pulse (duration of 1.2 ps; Fig .1b).

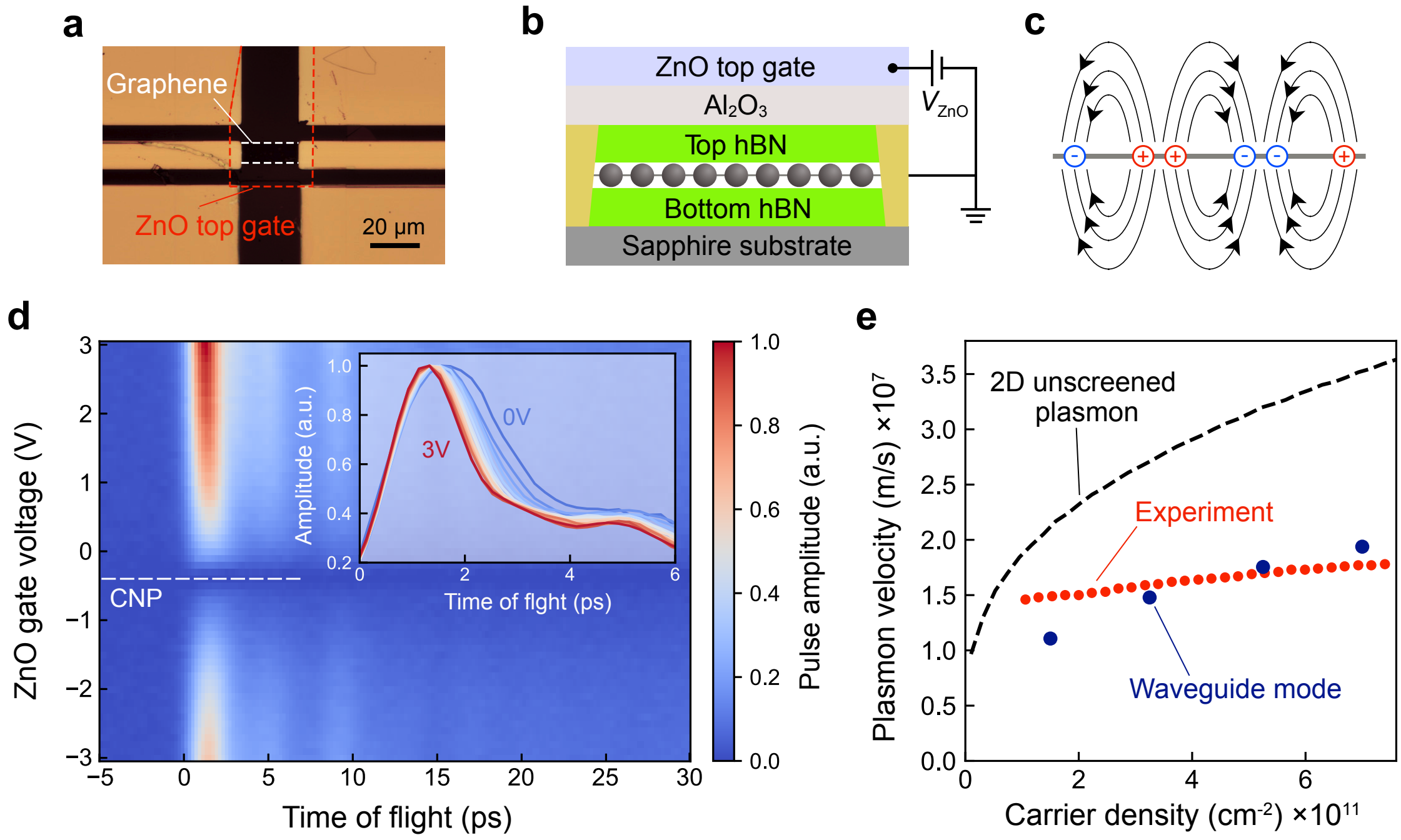

**Figure 2 | Unscreened plasmon transport in ZnO gate device. a,** Optical micrograph depicting the ZnO gate device. The geometric boundaries of the graphene and ZnO top gate are outlined by white and red dashed lines, respectively. **b**, Cross-sectional view of the sample. The transparency of the ZnO top gate to GHz and THz electric fields permits the formation of unscreened graphene plasmons. **c**, Schematic diagram of unscreened graphene plasmon with the associated charge and electromagnetic force. **d**, Time-domain waveforms for various ZnO gate biases. The pulse amplitude is normalized at the peak of the waveform obtained at $V_{ZnO}$ = 3.0 V. The horizontal dashed line indicates the charge neutrality point ($V_{CNP}$ = -0.35 V). The inset shows normalized time-domain waveforms around the peak position. The pulse duration is 2.0 ps (FWHM) at $V_{ZnO}$ = 3.0 V. **e**, Measured velocities of the plasmon (red circles, derived from **d**) are shown alongside the simulated velocities of the 2D unscreened plasmon (black dashed curve) and the waveguide mode plasmon (dark blue circles), plotted as a function of the carrier density. See Supplementary Section III for a detailed explanation of the simulation.

To quantitatively understand the changes in the plasmon velocity, we simulated the waveform of graphene plasmon wavepackets by using the dispersion of 2-dimensional (2D) unscreened graphene plasmons[1,2] $\omega_p \propto n^{\frac{1}{4}}k^{\frac{1}{2}}$ (Fig. 1c) and the measured waveform of the input electrical pulse (Fig. 1b) (refer to Supplementary Section III for the details of the simulation). Though the simulated wavepackets show qualitatively similar behaviour to the experimentally obtained ones (see the shortening of the pulse duration and increase in velocity with carrier density in Fig. S5), the measured velocity is approximately half of the simulated value and is much less sensitive to changes in the carrier density (Fig. 2e). These differences can be attributed to formation of a waveguide mode by lateral confinement in the graphene micro ribbon[47], where the wavelength of THz plasmons is predominantly determined by the lateral width of the ribbon. Indeed, the calculated phase velocity of the waveguide mode around the center wavelength (0.5 THz) shows quantitative agreement with the experiment (mode analysis on the graphene micro ribbon is detailed in Supplementary Section III). While the measured velocity is slightly less sensitive to carrier density changes compared to the calculated velocity of the waveguide mode, we speculate that this minor mismatch results from factors difficult to simulate, such as contact-induced doping[41] and quantum effects[48]. Our experimental data demonstrate that we can transfer ultrashort wavepackets of graphene plasmons in a waveguide. Furthermore, we can turn on/off the transfer by switching $V_{\mathrm{ZnO}}$ between the values for a large doping and the CNP.

**Acoustic graphene plasmons in an Au gate device**

Next, we examined the transport properties of graphene plasmons in a device with an Au top gate (Figs. 3a and b). The graphene is 8-μm wide and 26-μm long, and the length of the Au top gate is 12 μm. Owing to the screening effect induced by the Au top gate (Fig. 3c), the plasmon velocity is expected to be reduced significantly compared with the unscreened case, and the dispersion should be a linear relation[28,29] $\omega_p \propto k$ (Fig. 1c); because of this linearity, the screened plasmon is called an acoustic plasmon. In addition to the Au top gate, we incorporated a ZnO top gate to control the carrier density in the areas between the ohmic contacts and the Au gated regions. This allowed us to control the impedance matching between acoustic plasmons in graphene and the electrical pulse in the CPW (see Fig. S10 of the Supplementary Information). Throughout the measurements, graphene at the ZnO gated area was highly doped to minimize resistance ($V_{ZnO}$ = 3 V), ensuring effective conductive coupling. Figure 3d presents the time-domain waveform of the wavepacket as a function of the Au gate voltage ($V_{Au}$: -2.5 to 0.5 V, $n$: -5.5 to 7.1 × $10^{11}$ cm$^{-2}$). Note that the secondary peak around 10 ps originates from multiple reflections at both ends of the graphene and the PC switches, whereas our focus is on the first peak around 4 ps. The inset shows a normalized plot highlighting the peak shift near the CNP. Unlike the ZnO gate sample, which exhibits waveguide mode formation, the peak position shifts significantly with the gate voltage in the Au gate sample. Figure 3e shows the plasmon velocity calculated from the peak position and the propagation length determined by the Au gate length (12 μm). The velocity varies from 1.9 to 3.2 × $10^6$ m/s with doping. These values are close to the Fermi velocity (~1× $10^6$ m/s) and are one order of magnitude slower than those in the ZnO gate sample (Fig. 2e). The carrier density dependence of the velocity $v_{ac}$ above $n$ = 2 × $10^{11}$ cm$^{-2}$ is well reproduced by theory for the 2D acoustic plasmon[49] $v_{ac} \propto n^{\frac{1}{4}}$ (see Supplementary Section IV for details). Note that this agreement suggests that the waveguide mode is not formed in the Au gate sample. The absence of a waveguiding mode can be attributed to the fact that the characteristic size of the field distribution of the acoustic plasmon, determined by the separation of the graphene and Au gate (65 nm) and the plasmon wavelength (~2.5 μm at 1 THz), is much smaller than the width of the graphene micro ribbon (8 μm). On the other hand, the measured velocity deviates from the theoretical value for densities below $n$ = 2 × $10^{11}$ cm$^{-2}$ due to dissipation as discussed below.

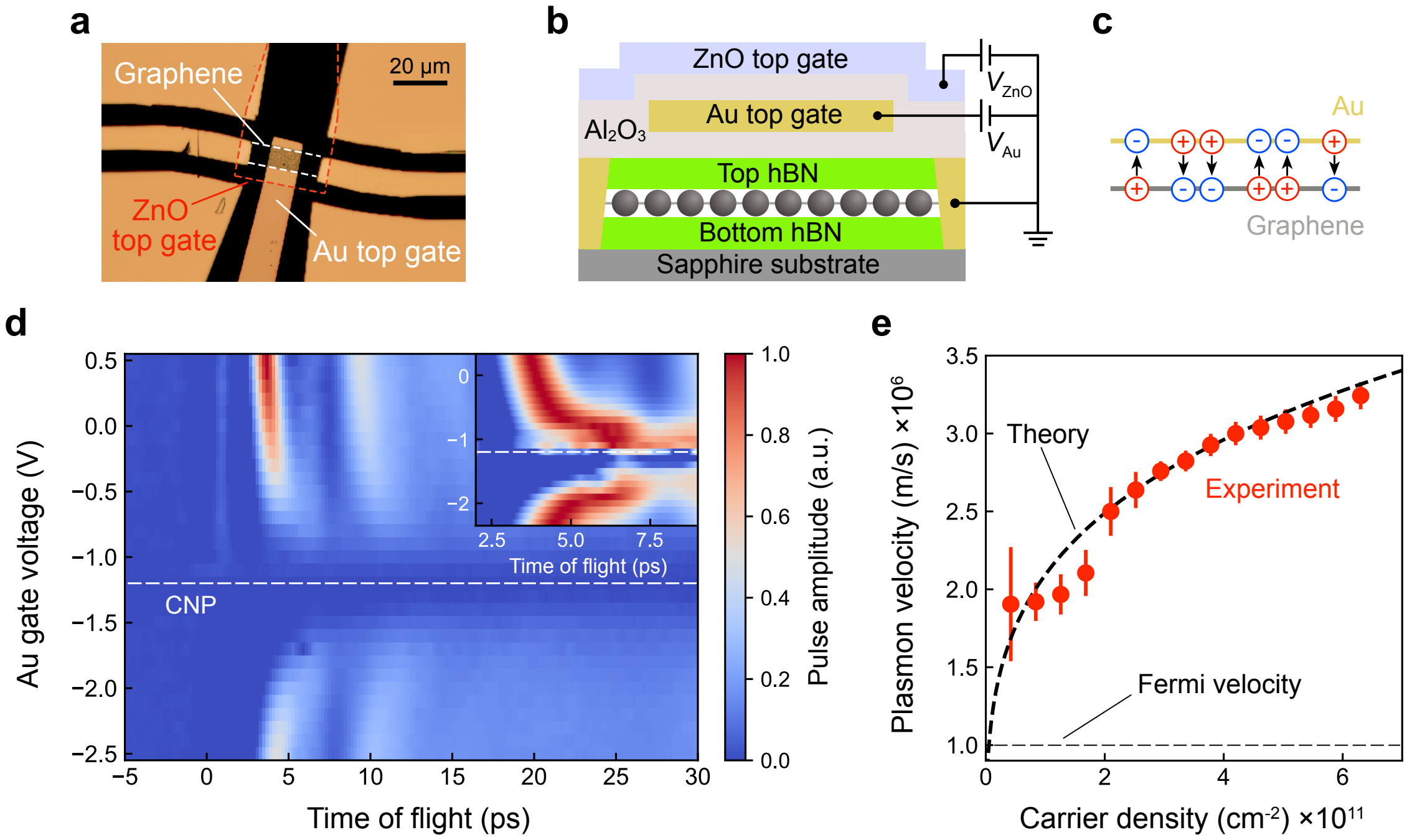

**Figure 3 | Acoustic plasmon transport in Au gate device. a,** Optical micrograph depicting the Au gate device. The geometric boundaries of the graphene and ZnO top gate are outlined by white and red dashed lines, respectively. **b**, Cross-sectional view of the sample. The Au gate is used to induce screening for forming acoustic plasmons. The ZnO gate bias is fixed at a large value, $V_{\mathrm{ZnO}}$ = 3 V, to maximize the coupling strength between the electrical pulse in the CPW and plasmons in graphene. **c**, Schematic diagram of acoustic plasmon with the associated charge and electromagnetic force. **d**, Time-domain waveform for various Au gate biases. The pulse amplitude is normalized at the peak of the waveform obtained at $V_{\mathrm{Au}}$ = 0.5 V. The horizontal dashed line indicates the charge neutrality point ($V_{\mathrm{CNP}}$ = -1.2 V). The inset shows a normalized plot around the peak position. **e**, Measured (red circles, derived from **d**) and calculated plasmon (black dashed curve) velocities as a function of carrier density. Error bars are due to the uncertainty of determining the peak positions.

**Coherent and incoherent transport of ultrashort wave packet**

For a comprehensive understanding of the transport properties of the acoustic plasmon, let us analyse the time-domain waveform in the $n$-doped region (Fig. 4a). Remarkably, the measured waveform at $V_{Au}$ = 1.5 V ($n$ = 11.3 × $10^{11}$ cm$^{-2}$) has a pulse duration of 1.2 ps (FWHM), which, to the best of our knowledge, represents the shortest electrically generated plasmon wavepacket observed in any material: this duration is substantially shorter than ~25 ps for graphene edge magnetoplasmons[36] and ~9 ps for 2D plasmons in a GaAs/AlGaAs heterostructure[40]. The wavepacket propagates with a velocity of 3.3 × $10^6$ m/s and is directionally confined to 4.0 μm. Together with the vertical confinement of 65 nm (30 nm $Al_2O_3$ layer and 35 nm hBN layer) due to the screening effect (calculated field profile is shown in Fig. S4) and the lateral confinement of 8 μm due to the width of the graphene, this means that the plasmon wavepacket is confined within a $2.1 \times 10^{-18}$ m$^3$ volume. The volume normalized by free space wavelength (calculated from the center frequency of 0.28 THz obtained from the inset of Fig. 1b and the speed of light of $3 \times 10^8$ m/s) is $5.2 \times 10^{-9}$. This extremely tight confinement would be beneficial for constructing various graphene plasmonic nanocircuits. It is worth noting that the width of 1.2 ps for the plasmon wavepacket is the same as that of the input electrical pulse, which is consistent with the linear dispersion of the acoustic plasmon. As the carrier density is decreased, the pulse duration becomes longer than 20 ps at $V_{Au}$ = -1.0 V ($n = 0.8 \times 10^{11}$ cm$^{-2}$). The range of carrier densities exhibiting this pulse broadening corresponds to the range where the velocity deviates from the theoretical acoustic plasmon velocity (Fig. 3e).

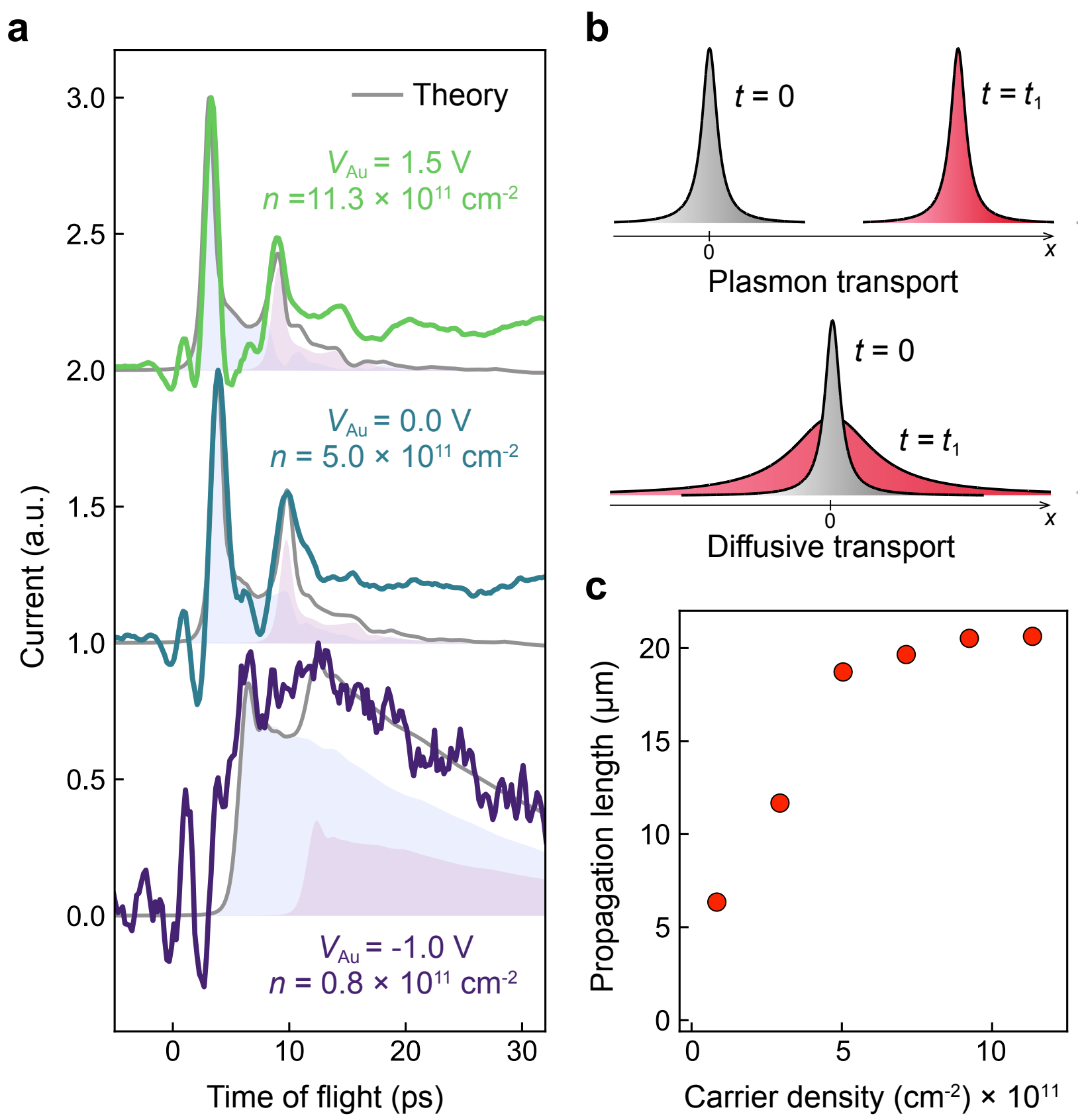

**Figure 4 | Waveform comparison between experiment and theory. a**, Experimental (colored traces) and simulated (gray traces) normalized time-domain waveforms of wavepackets for three Au gate biases after 12 μm propagation. The simulation includes the initial (blue area) and the first echo (purple area) pulses (see Supplementary Section IV for details). Each waveform is offset by unity for clarity. **b**, Schematic diagram of wavepacket transport in real space in the different transport regimes. **c**, Simulated propagation length of acoustic plasmon wavepackets as a function of the carrier density obtained by the telegrapher's equation[49].

The evolution of the waveform with carrier density can be explained by a simulation solving the telegrapher's equation for an *RLC* distributed constant circuit[35,49], where *R*, *L*, and *C* correspond to the resistivity of graphene, graphene kinetic inductance, and capacitance between graphene and the Au gate (for details, refer to Supplementary Section IV). The initial and the first echo pulse from the PC switches are included in the simulation. The simulation accurately reproduces the experimental waveform, including the pulse width and the peak position, for the whole carrier density range. The solution to the telegrapher's equation[49] suggests that crossover from coherent plasmonic transport to incoherent diffusive transport (Fig. 4b) occurs as the carrier density is decreased to the CNP. In the heavily doped region ($V_{Au}$ = 1.5 V), the wavepacket traverses the graphene in a coherent manner as acoustic plasmons. Hence, the pulse duration remains identical to the input electrical pulse. Conversely, near the CNP ($V_{Au}$ = -1.0 V), where the resistivity is high, the acoustic plasmon dissipates before traveling 12 μm due to the shorter lifetime, leading to a broader pulse through incoherent diffusive transport. Since diffusive transport is much slower than plasmon transport, the peak of the waveform shifts with a longer time delay when diffusive transport becomes dominant near the CNP.

With this model, we can deduce the propagation length of acoustic plasmons (Fig. 4c). At the highest doping level, the propagation length $l_p$ reaches 21 μm. With a pulse width of 4.0 μm, more than five wavepackets can be transferred before decoherence. Notably, this exceeds the previously reported value of 2.1, obtained with a plasmonic wavelength of 222 nm and $l_p$ = 470 nm, which was measured using near-field scanning optical microscopy on CVD graphene at room temperature[32]. Since our model predicts that the propagation length scales inversely with resistivity, $l_p \propto 1/R$, further improvements to the carrier mobility through optimization of the device fabrication process could lead to an even greater $l_p$.

**Conversion efficiency of graphene plasmon excitation**

Finally, let us discuss the efficiency of conversion between the electrical pulse and the graphene plasmon wavepacket. We expected that the electrical excitation used in this study would have high conversion efficiency as it bypasses some of the restrictions associated with optical excitations, such as substantial momentum mismatches[16–33] and graphene's small nonlinear coefficient[50]. To demonstrate this, we plotted the time-domain waveform of the unscreened and acoustic plasmon wavepackets together with the input electrical pulse with the amplitude normalized relative to the input electrical pulse (Fig. 5a); for the normalization, the responsivity of the PC switches was adjusted (see Supplementary Section I-b). Figures 5b and c show the transmittance, defined as the ratio of the peak amplitudes $I_{out}/I_{input}$ of the input and output electrical pulses. This transmittance value is affected by the following three processes: (1) conversion of the input electrical pulse into the graphene plasmon wavepacket ($I_{input} \rightarrow I_{plasmon}$); (2) attenuation of the plasmon wavepacket as it moves within the graphene; (3) conversion of the plasmon wavepacket into the output electrical pulse ($I_{plasmon} \rightarrow I_{output}$). Given that processes (1) and (3) take place at CPW-graphene interfaces under identical conditions ($I_{plasmon}/I_{input} = I_{output}/I_{plasmon}$), the transmittance provides a lower bound for the energy conversion efficiency, $\eta = (I_{plasmon}/I_{input})^2$. Namely, $I_{out}/I_{input} = (I_{plasmon}/I_{input})^2\zeta = \eta\zeta$, where $\zeta (\leq 1)$ is the plasmon attenuation in process (2). For unscreened plasmons, $\eta$ can exceed 0.35 at high carrier densities. For acoustic plasmons, since the propagation length is 21 μm, $\zeta \sim 0.56$ and $\eta$ can be as large as 0.03. These values are orders of magnitude higher than the previously reported $\eta = 6 \times 10^{-5}$ for optical excitation of propagating THz graphene plasmons[50]. These unusually high conversion efficiencies come from the fact that energy losses due to momentum mismatches are absent for the conversion between the electrical pulse and the plasmon wavepacket. Instead, we suggest that the high-frequency impedance mismatch between the CPW and the graphene transistor is a limiting factor in our devices. In the Au gate device, the large difference in velocity between the electrical pulse in the CPW (~$1 \times 10^8$ m/s) and the acoustic plasmon (~$3 \times 10^6$ m/s) and also the presence of the Au gate induce a large impedance mismatch. In the ZnO gate device, on the other hand, the smaller velocity difference due to the higher plasmon velocity (~$1 \times 10^7$ m/s) and the transparent ZnO gate in the THz range result in a smaller impedance mismatch and thus a higher conversion efficiency. It is important to note that the frequency component of the plasmon wavepacket transferred from the input electrical pulse covers a broad frequency range. Our ability to produce propagating plasmons over a wide frequency range with significant efficiency will be pivotal to the development of graphene plasmonic circuitry[5,7–13]. We foresee the potential for further enhancement of the conversion efficiency, for instance, by constructing subwavelength structures that reduce the impedance mismatch between the CPW and the graphene transistor.

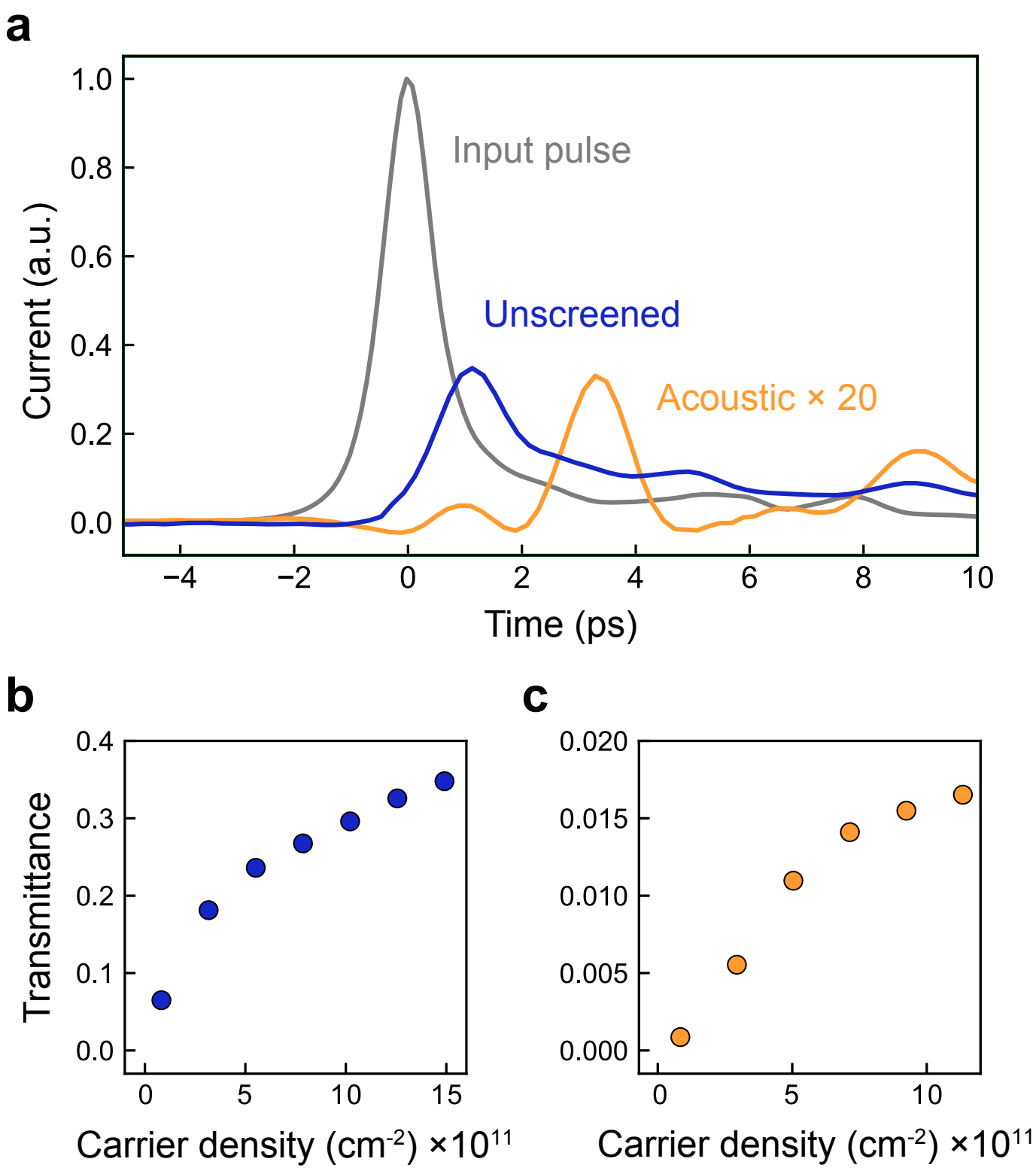

**Figure 5 | Conversion efficiency for graphene plasmons.** **a**, Time-domain waveform of input electrical pulse together with the unscreened (ZnO gate: $n$ = 15.0 × $10^{11}$ $cm^{-2}$) and acoustic (Au gate: $n$ = 11.3 × $10^{11}$ $cm^{-2}$) graphene plasmon wavepackets. The amplitude of the wavepacket is normalized to the peak amplitude of the input pulse. The amplitude of the acoustic plasmon is magnified by a factor of 20 for clarity. **b**,**c**, Transmittance of the unscreened (**b**) and acoustic (**c**) graphene plasmon as a function of the carrier density. Transmittance is the ratio of the peak amplitude to that of the input pulse.

## Conclusions

We have reported the on-chip generation, manipulation, and detection of ultrashort graphene plasmon wavepackets using THz electronics. An ultrashort electrical pulse is injected into a graphene microribbon via an ohmic contact and the transport properties of unscreened and acoustic graphene plasmon wavepackets in the THz regime can be studied by manipulation of the dielectric environment with ZnO and Au top gates. The unscreened plasmons in the ZnO gate device have a velocity on the order of $10^7$ m/s, with waveguiding modes formed due to the lateral confinement of the graphene microribbon, while the velocity of the acoustic plasmons in the Au gate device drop by an order of magnitude to $10^6$ m/s. At high carrier densities, the injected wavepackets propagate coherently as acoustic plasmons with a minimum pulse width of 1.2 ps. Conversely, near the CNP they diffuse incoherently, resulting in a broader pulse. Notably, the shortest wavepacket is tightly confined within a volume, $2.1 \times 10^{-18}$ $m^3$, making acoustic plasmons an ideal candidate for constructing graphene plasmonic nanocircuits. The propagation length of acoustic plasmons is estimated to be 21 μm by waveform analysis.

A key advantage of our excitation scheme using spatiotemporally localized electrical pulses is the high conversion efficiency between electrical pulses and plasmon wavepackets, reaching $\eta$ = 0.35 and 0.03 for unscreened and acoustic plasmons, respectively. This is a large improvement over that achieved by optical excitation of propagating graphene plasmon wavepackets. By overcoming the limitations of conventional optical techniques, our method could serve as a guide for constructing graphene plasmonic circuits[5,7–13] and integrated devices[51,52] for information transfer and processing. The ability to excite and measure electrical pulses as short as 1.2 ps could also be used in the study and manipulation of the transport properties of various quantum materials and circuits beyond graphene, such as van der Waals heterostructures[3–5] and quantum nanoelectronic devices[53–56]. The use of THz electronics to shrink pulse duration by more than one order of magnitude compared to conventional GHz electronics[55] could be of value in fields where the size of plasmon or single-electron wavepackets[53–56] are substantially smaller than the material size and coherence length.

## Methods

### Device fabrication

Photoconductive switches were prepared using an LT-GaAs wafer supplied by BATOP, GmbH. The wafer consisted of a 2.6-μm-thick LT-GaAs surface layer (grown at 300 °C) and a 500-nm-thick $Al_{0.9}Ga_{0.1}As$ sacrificial layer on a semi-insulating GaAs substrate. After the LT-GaAs layer was etched into 100 μm × 100 μm squares by using a citric acid solution, the sacrificial layer was dissolved in hydrochloric acid. The etched LT-GaAs chips on the GaAs substrate were transferred to a sapphire substrate by using a thermoplastic methacrylate copolymer (Elvacite 2552C, Lucite International) as an adhesive[57]. The remaining Elvacite on the sapphire substrate was removed with citric acid.

Graphene was prepared via mechanical exfoliation of natural graphite on silica (285 nm)/doped silicon substrates. Monolayer graphene was identified via optical contrast under a microscope. Using a separately exfoliated hBN flake, graphene was picked up and transferred onto the sapphire substrate by using the dry-transfer technique with polydimethylsiloxane and polycarbonate[45]. The thickness of the hBN layer on top of the graphene was 72 nm for the ZnO gate device and 35 nm for the Au gate device. The carrier mobility within the graphene, measured at 4 K, was 250,000 $cm^2V^{-1}s^{-1}$ for the ZnO gate device and 73,000 $cm^2V^{-1}s^{-1}$ for the Au gate device (refer to Supplementary Section VII for the details).

The graphene was patterned by reactive ion etching, and the Ti/Au waveguide structure with side contacts to graphene was deposited by sputtering. The whole surface was covered with a 30-nm-thick alumina ($Al_2O_3$) insulating layer grown by atomic layer deposition. For the Au gate device, Ti/Au (10/100 nm) was patterned on the $Al_2O_3$ layer by using photolithography, deposition, and lift-off processes. Then, for both devices, a 20-nm-thick ZnO gate was grown and patterned by using atomic layer deposition at 200°C. The ZnO gate was protected by another $Al_2O_3$ layer deposited on top. Finally, in order to make electrical contact with the waveguide, Miroposit 351 developer was used to remove the $Al_2O_3$ on the bonding pads. We noted that the photolithography processes typically introduce fabrication errors on the order of a few hundred nanometers.

**On-chip THz spectroscopy measurements**

A femtosecond laser (Monaco, Coherent, Ltd.) was used as the light source (1,035 nm, 280 fs, 16.8 MHz). Two orthogonally polarized pump and probe beams were combined using a polarization beam splitter and aligned with a slight displacement in order to focus them onto the two LT-GaAs PC switches by using an objective lens[58] for generation and detection of ultrashort electrical currents. The pump power was 20 mW for both PC switches, and a 30 V bias was applied to the injection PC switch. An optical chopper modulated the pump beam at a few hundred hertz for lock-in detection of the terahertz current.

Measured signals include crosstalk. The crosstalk component can be determined at the CNP, where plasmons cannot propagate. Since the crosstalk is independent of the graphene carrier density, plasmon signals free from the crosstalk can be obtained by subtracting the signal at the CNP from those at finite carrier densities. This method is applied to both ZnO and Au gated devices.

## Data availability

The datasets generated during and/or analysed during the current study are available from the corresponding author on reasonable request.

## Acknowledgements

The authors thank L. Pugliese for fruitful discussions and H. Murofushi for technical support. This work was supported by JSPS KAKENHI Grant Numbers 21H05233, 23H02052, JP22H00112 and JP24H00828. K.W. and T.T. acknowledge support from World Premier International Research Center Initiative (WPI), MEXT, Japan.

## Author contributions

K.Y. and N.K. conceived the experiment. K.Y. designed and built the optical setup and conducted finite element calculation. K.Y. and G.B. performed the measurement, analysed the data, and conducted the unscreened plasmon simulation. K. S. simulated the acoustic plasmon wavepacket. K.Y., G.B. and N.K. designed the THz circuits with support from T.W. and M.H. G. B., T.W., S.S. and N.K. fabricated the devices. K.W. and T.T. contributed the hBN material. K.Y. and N.K. wrote the paper, with input from all authors.

## Competing financial interests

The authors declare no competing financial interests.

**Supplementary Information for**

# On-chip transfer of ultrashort graphene plasmon wavepackets using terahertz electronics

Katsumasa Yoshioka[1,*], Guillaume Bernard[1], Taro Wakamura[1], Masayuki Hashisaka[1], Ken-ichi Sasaki[1], Satoshi Sasaki[1], Kenji Watanabe[2], Takashi Taniguchi[3], and Norio Kumada[1]

[1]NTT Basic Research Laboratories, NTT Corporation, Atsugi, 243-0198, Japan

[2]Research Center for Electronic and Optical Materials, National Institute for Materials Science, Tsukuba, Japan

[3]Research Center for Materials Nanoarchitectonics, National Institute for Materials Science, Tsukuba, Japan

*e-mail: katsumasa.yoshioka@ntt.com

## I-a. Determining time origin and propagation velocity

In our on-chip THz spectroscopy, the peak of the THz current signals the moment when the travel time of the generated wavepacket between the two photoconductive (PC) switches matches the time delay of the probe beam from the pump beam (Fig. S1a). The point at which the pump and probe beams arrive simultaneously is the midpoint of the two wavepackets and it is determined by switching the pump and probe PC switches (Fig. S1b). The propagation velocity in the coplanar waveguide (CPW) is then calculated for a device without graphene by dividing the distance between Switch A and B by the time delay at the current peak. The calculated propagation velocity is $v_{\mathrm{CPW}} = 1.20 \times 10^8$ m/s, in good agreement with the literature value[1]. The time used in the main manuscript is obtained by eliminating the propagation time in the CPW calculated using $v_{CPW}$ and the length of the CPW from the full time delay.

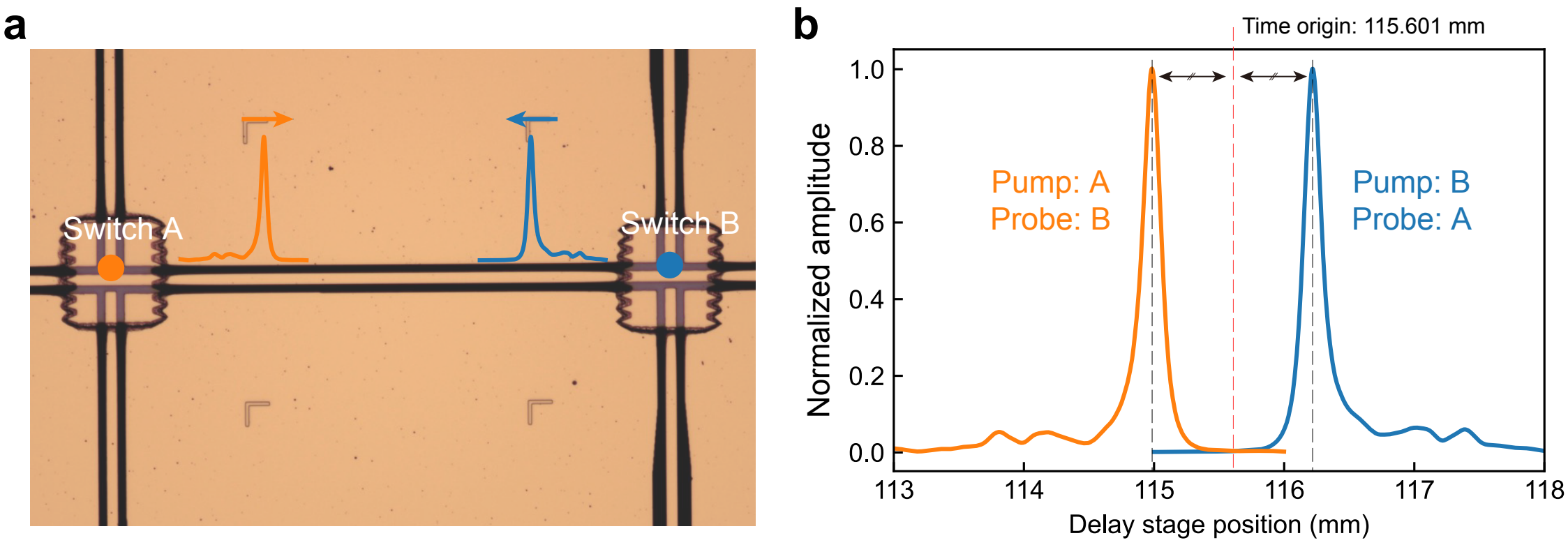

**Figure S1 | Pump switching experiment in a coplanar waveguide. a**, Optical micrograph of a CPW. The distance between PC switches is 485 μm. **b**, Normalized waveform obtained by pumping different PC switches A and B. Corresponding pump spot positions are depicted with coloured circles in **a**.

**I-b. Calibration of PC switch sensitivity**

Our PC switches were fabricated from the same low-temperature-grown gallium arsenide (LT-GaAs) wafer, ensuring a uniform response time across all devices. However, as shown in Fig. S2a, there are variations in responsivity to incident femtosecond laser pulses between different PC switches. (Switches A and B in Fig. S2a are the same as those shown in Fig. S1a). This difference in I-V characteristics originates from minor variations in the contact resistance of the readout electrode and the gap length of the PC switches. The impact of these differences on the measured THz currents is illustrated in Fig. S2b. Here, the excitation conditions, such as the DC bias voltage (30 V) applied to the pump switch, as well as the optical spot size and power, were kept constant. As anticipated, the dynamic response between the two waveforms remains the same. However, the signal amplitude of the configuration with Pump: A, Probe: B (orange) exhibits a signal 50% smaller than that of the other configuration (blue). This confirms that the pump and probe switches have different effects on the measured THz signals.

For the pump switch, the generated current amplitude is determined by the DC current amplitude at 30 V, as the amount of integrated DC current is directly related to the peak amplitude of the generated THz current. For the probe switch, the readout sensitivity is determined by the slope of the I-V curve in the linear regime (-5 to 5 V). This is because the amplitude of the probe current is determined by the driven photocurrent due to the small voltage of the generated THz current propagating through the CPW. By taking these different I-V characteristics into account, we can calibrate the current amplitude of the measured THz current as shown in Fig. S2b (black dashed curve). This calibration verifies that, after determining the light-excited I-V curves for the pump and probe PC switches, the amplitude of the THz current can be compared across different devices, as in Fig. 5a-c of the main text.

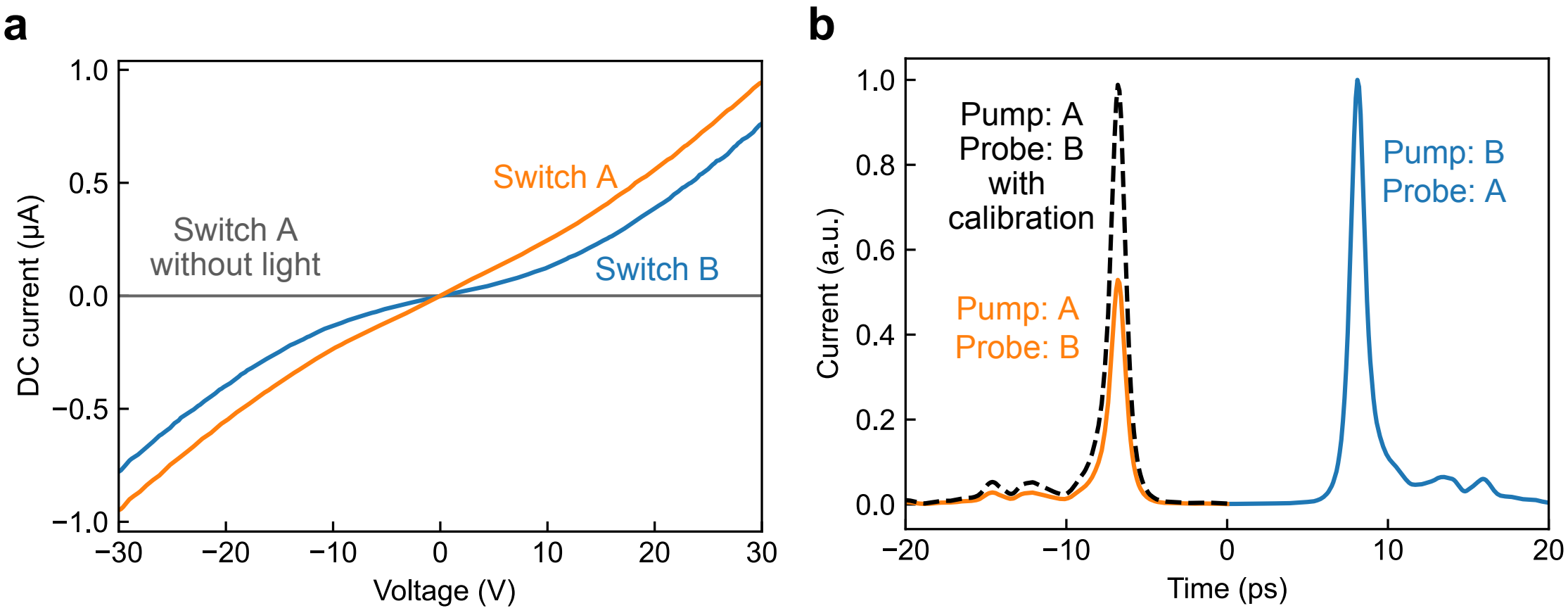

**Figure S2 | Calibrating the amplitude of the measured THz current. a**, I-V characteristic of PC switches A and B upon femtosecond laser excitation. Both switches exhibit insulating behaviour without laser excitation under the experimental condition at 4 K. **b**, Measured THz current. The amplitude of the THz current can be calibrated using the I-V characteristics of the pump and probe PC switches.

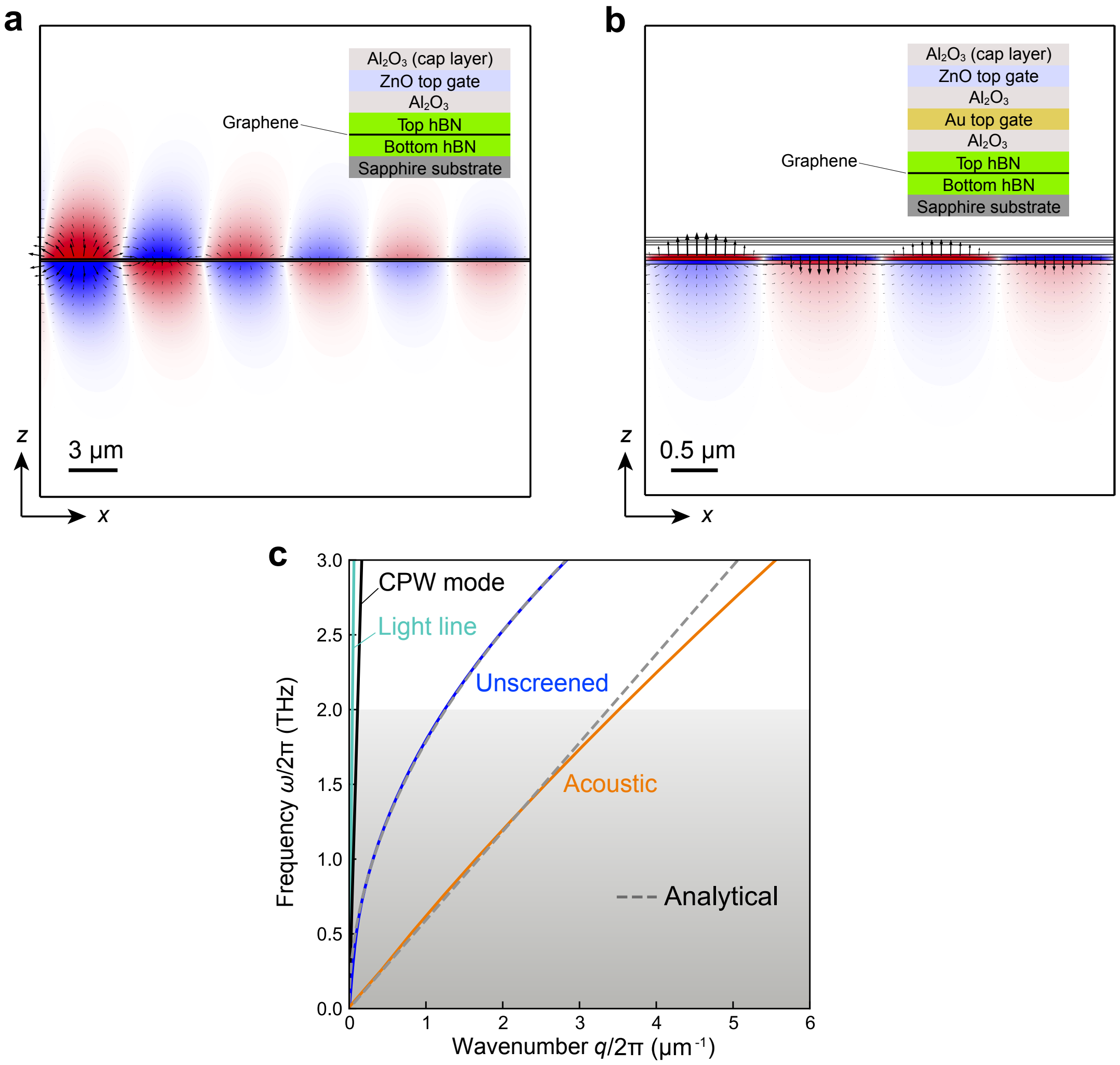

**Figure S3 | Finite element calculation of graphene plasmon. a,b**, Simulation of unscreened (**a**) and acoustic (**b**) plasmon propagation at a frequency of 1 THz for a carrier density of $2.5 \times 10^{11}$ $cm^{-2}$ in uniform graphene. The inset shows layer structures incorporated in the calculation. **c**, Dispersion curve of graphene plasmon for the same carrier density in **a** and **b**. The blue and orange curves represent the dispersion numerically derived from the simulation in **a** and **b** across varying frequencies. The grey dashed curves corresponds to the dispersion curve derived from the analytical theories[2,3], used in our wavepacket analysis (the dashed curve for unscreened plasmon overlaps with the blue solid curve). CPW mode corresponds to the even mode excited in the CPW (Fig. S7). The grey shaded area indicates the range of frequencies excited by our input electrical pulse.

## II. Finite element analysis of graphene plasmon

For precise estimation of the dielectric environment and its impact on the velocity of the plasmon wavepacket, we employed finite element calculations using COMSOL software. These calculations were performed on uniform graphene (without a micro-ribbon structure), incorporating the exact layer structure of our ZnO and Au gated devices as depicted in Fig. S3a and b. Figure S3c presents the dispersion obtained by these simulations, plotted alongside the analytical theories[2,3] that were utilized in our wavepacket simulations detailed in Sections III and IV. The strong correlation between the finite element calculations and analytical theories substantiates the validity of our wavepacket analysis. Additionally, Fig. S4 provides a detailed view of the normalized electric field of the acoustic plasmon, derived from the simulation shown in Fig. S3b. This calculation reveals that most of the electric field is confined within the hBN and $Al_2O_3$ layers between the graphene and the Au top gate, with minimal extension into the sapphire substrate.

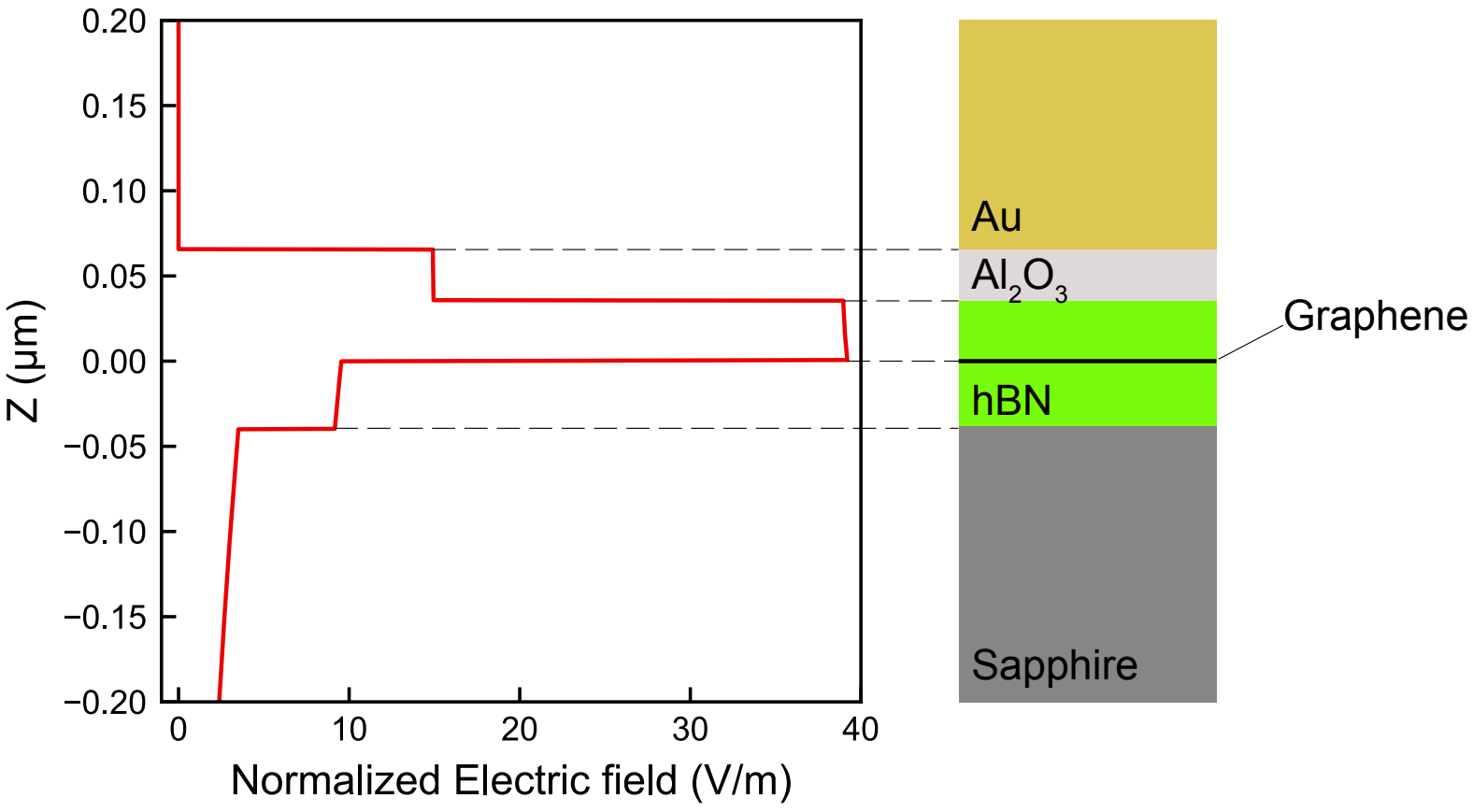

**Figure S4 | Calculated field profile of acoustic plasmon.** A close-up view of the normalized electric field, derived from the simulation in Fig.S3b. Schematic of the corresponding sample structure is provided for reference on the right panel.

### III. Simulation of unscreened graphene plasmon wavepackets

Figure S5a shows the calculated dispersions of unscreened graphene plasmons in the THz regime obtained by considering them to be a 2-dimensional (2D) Dirac plasma[2]. The background dielectric constant was determined by fitting the calculated dispersion curve to the one obtained from the finite element simulation, as shown in Fig. S3c. The slope of the dispersion gets steeper as the carrier density increases. With these dispersion relations and measured waveform of the input THz electrical pulse (identical to the waveform displayed in Fig. 1b of the main text), we simulated the waveform of unscreened graphene plasmon wavepackets after 23-μm propagation (Fig. S5b). By increasing the carrier density, the velocity increases, and the pulse duration shortens. This trend qualitatively aligns with the experimental data presented in the inset of Fig. 2d in the main text. However, as discussed in the main text, the experimental velocity is approximately half of the simulated value due to the formation of a plasmonic waveguide in the graphene micro ribbon[4].

To estimate the velocity of this waveguide mode, we conducted a mode analysis,

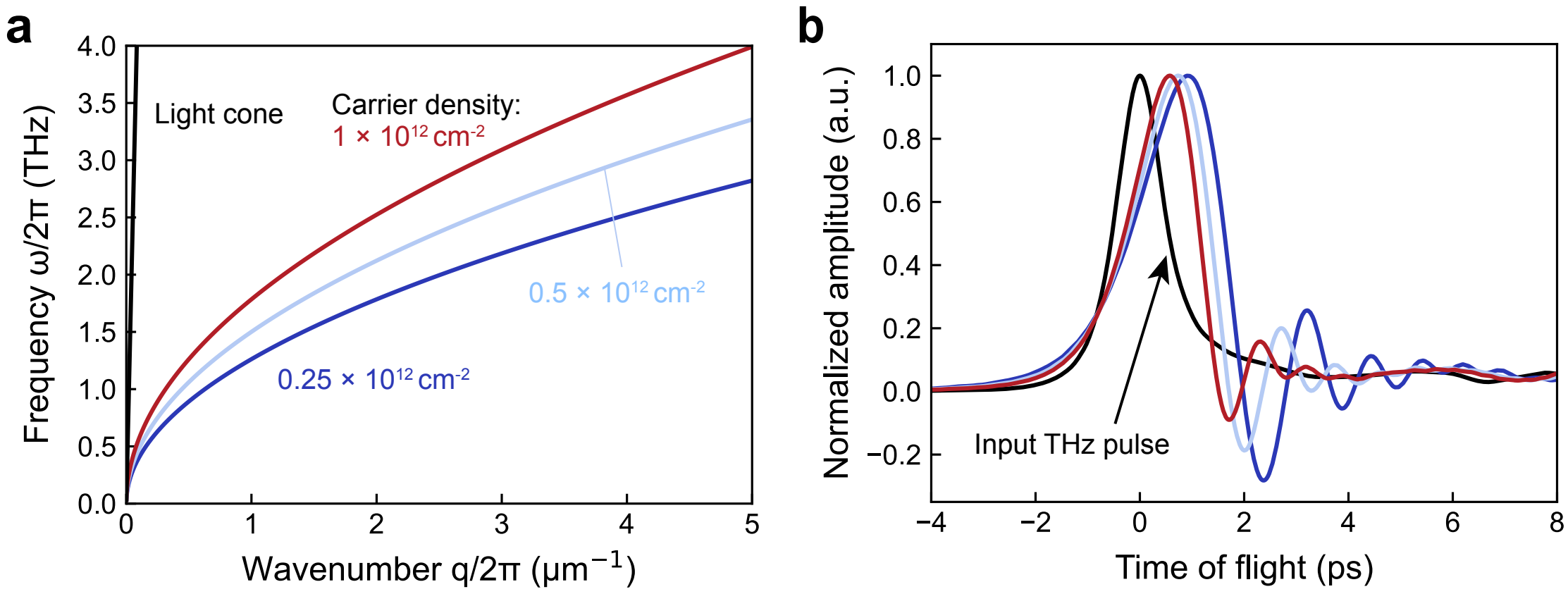

**Figure S5 | Simulated waveforms of unscreened graphene plasmon. a**, Calculated dispersions of unscreened graphene plasmon for different carrier densities. **b**, Simulated waveforms of unscreened graphene plasmon. The colour of the waveforms corresponds to the dispersion used for the simulation in **a**. Simulated plasmon velocity (Fig. 2e in the main text) is calculated for the positions of the peaks of these wavepackets.

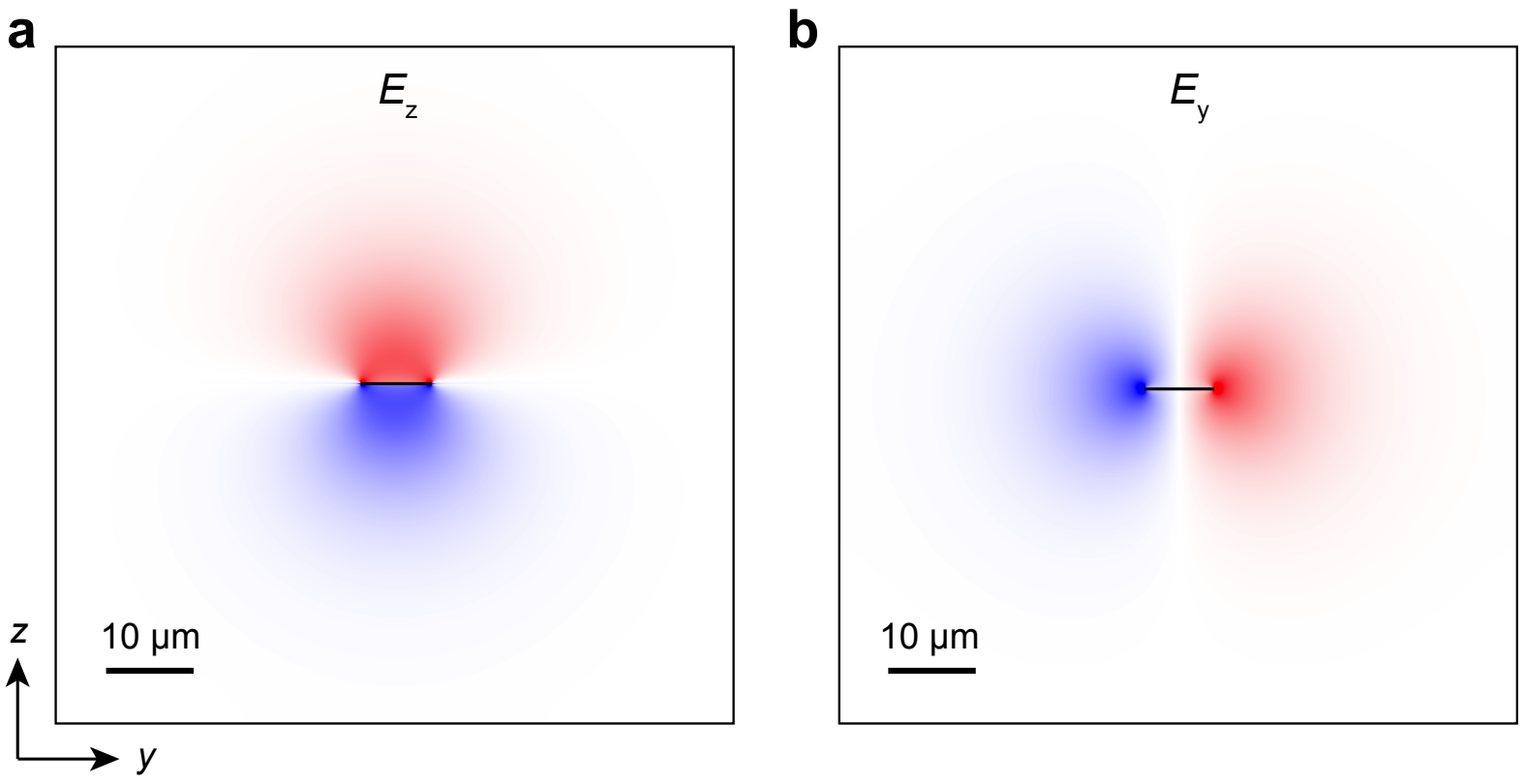

**Figure S6 | Mode analysis of unscreened graphene plasmon. a,b**, Color plots depicting the z (**a**) and y (**b**) components of the electric field at a frequency of 0.5 THz for a carrier density of $5.25 \times 10^{11}$ cm$^{-2}$. These plots illustrate the fundamental plasmonic mode within the device, with an effective refractive index of 17.1.

presenting the fundamental mode in Fig. S6a and b. For this analysis, we selected a fixed frequency of 0.5 THz, close to the center frequency of the input electrical pulse. Utilizing the effective refractive index obtained, we calculated the phase velocity of the waveguide mode plasmon, as depicted in Fig. 2e of the main text. This shows a good agreement with experimental results, displaying a slower velocity than predicted by 2D unscreened plasmon theory, as expected[4]. Additionally, we conducted mode-dependent excitation of graphene plasmons by changing the bias polarity applied to the PC switch as depicted in Fig. S7. It demonstrates that graphene plasmon is generated only when the symmetric even mode is excited in the CPW, which has the same field distribution with the fundamental plasmonic mode (Fig. S6a and b). This selectivity suggests that higher-order plasmonic modes are not supported in our THz circuit.

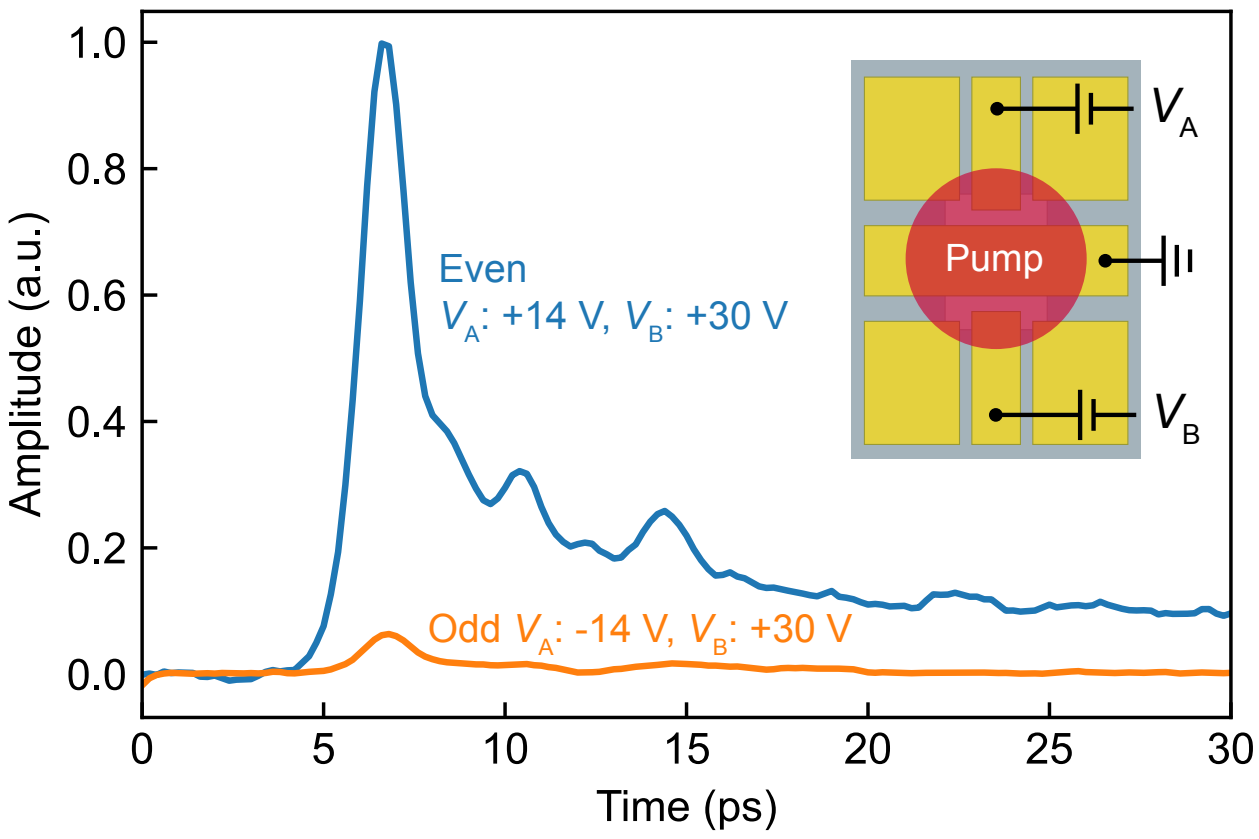

**Figure S7 | Selective excitation of even and odd modes.** Time domain waveforms of unscreened graphene plasmon in the ZnO gate device ($V_{ZnO}$ = 3.0 V) under different excitation modes. The inset depicts the experimental setup for selectively exciting even or odd modes. The bias voltage applied to the upper PC switch ($V_A$) is set lower than the other ($V_B$) to equalize the THz current amplitude, calibrated by measuring light-excited I-V curves for both PC switches.

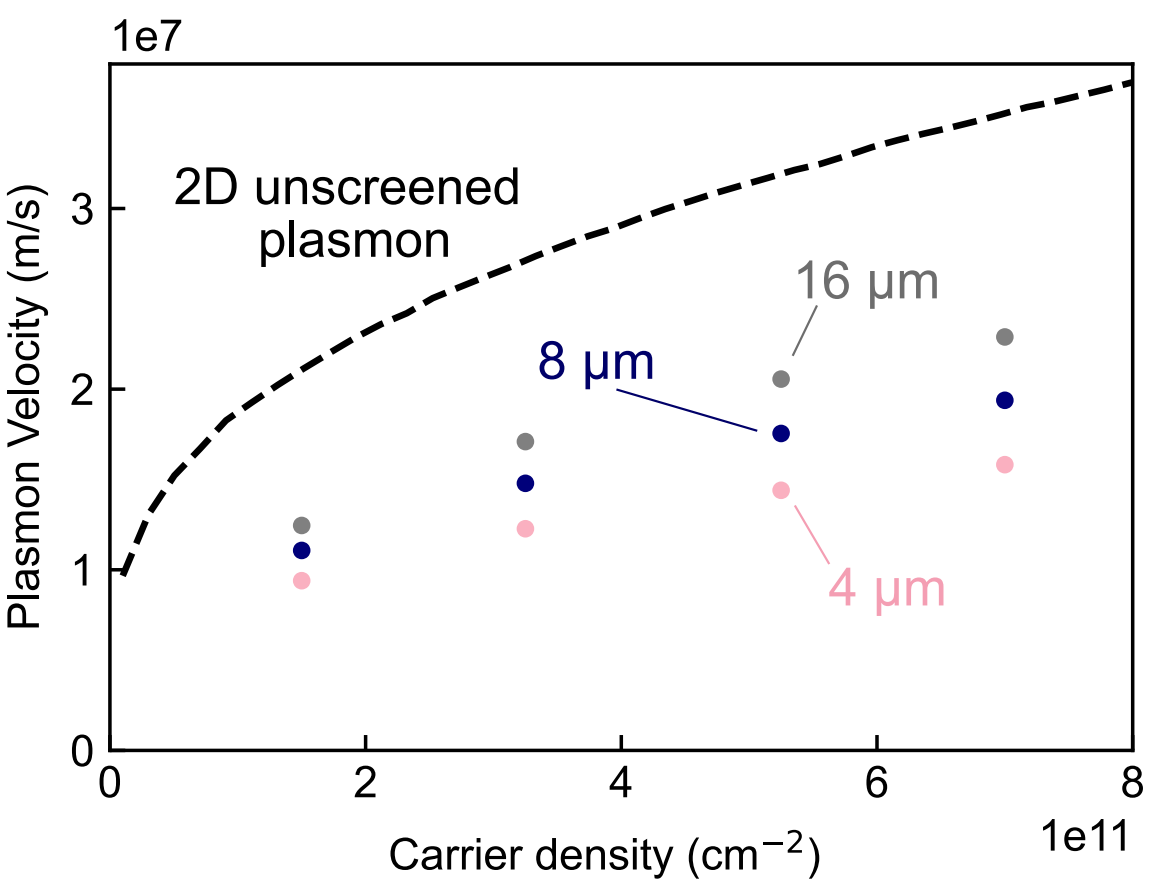

**Figure S8 | Plasmon velocity variation with graphene width.** The coloured circles represent the unscreened plasmon velocities corresponding to various widths of graphene obtained by the mode analysis.

To further elucidate the formation of the waveguide mode, we examined the dependence of the graphene width on the waveguide properties, as illustrated in Fig. S8. We found that a narrower graphene width results in a reduced velocity compared to the 2D plasmon, further substantiating the formation of the waveguide mode in our ZnO gated device.

## IV. Simulation of acoustic plasmon wavepackets

The velocity of acoustic plasmons in graphene shown in Fig. 3e of the main text is theoretically obtained by neglecting the finite resistance of graphene,[3]

$$v_{ac} = 2e\left(\frac{v_F}{C\hbar}\right)^{\frac{1}{2}}(\pi n)^{\frac{1}{4}}, \qquad \text{(S1)}$$

where $e$, $v_F$, $C$, $n$ correspond to the electron charge, Fermi velocity, capacitance between graphene and the Au gate, and carrier density. This implies that the $v_{ac}$ could be even slower in cases of thinner $Al_2O_3$ and hBN due to the enlarged $C$. Experimental results are well reproduced by Eq. (S1) with $C = 9.7 \times 10^{-4}$ Fm$^{-2}$, which is in good agreement with the geometrical capacitance, $C_d = 6.7 \pm 1.8 \times 10^{-4}$ Fm$^{-2}$.

The waveform of acoustic plasmons in an actual graphene device with finite resistance can be obtained by using a distributed constant circuit model[5]. The response of charge carriers in graphene to a high-frequency electric field is characterized by the impedance $Z_{\mathrm{graphene}} = R + i\omega L$, where $L$ is the kinetic inductance arising from the inertia of the charge carriers. Since the carrier dynamics depends on $n$, $L$ can be modified by changing $n$ as $L = \hbar/4v_F e^2\sqrt{\pi n}$. Including the capacitive coupling to the metal gate $C$, the wave equation in the propagation direction $x$ can be written in the form of the telegrapher's equation,

$$\left(\frac{\partial^2}{\partial t^2} - \frac{1}{LC}\frac{\partial^2}{\partial x^2} + \frac{R}{L}\frac{\partial}{\partial t}\right)V(x,t) = 0, \tag{S2}$$

where $V$ is the potential induced by excess carriers in graphene. This gives the plasmon dispersion,

$$\omega = \sqrt{\frac{k^2}{LC} - \frac{R^2}{4L^2}}. \tag{S3}$$

Note that Eq. (S3) gives the velocity $v = 1/\sqrt{LC}$ for $R = 0$, which corresponds to Eq. (S1). The corresponding dispersion is illustrated in Fig. S3c (represented by a gray dashed line), alongside the dispersion derived from the finite element calculation. The close agreement between these two dispersions confirms the validity of our acoustic plasmon analysis. Furthermore, from the solution to Eq. (S2), the propagation length $l_{\mathrm{p}}$ of the acoustic plasmon can be determined as

$$l_p = \left(\frac{R}{2}\sqrt{\frac{C}{L}}\right)^{-1}. \tag{S4}$$

To compare with the experimental waveform (Fig. 4a in the main text), we performed a numerical simulation of the wavepacket by discretizing the telegrapher's equation, Eq. (S2), with a square lattice network in graphene. The square lattice network consists of *RLC* circuit elements shown in the inset of Fig. S9a. At the *i*-th site of the network, the potential $V_i(t)$ is defined as a function of time $t$. The Au gate is taken into account by $C$ at each site. The link between nearest-neighbor sites (separation is $\Delta x = 0.5$ μm) consists of $L$ and $R$ connected in a series. The current from site $j$ to $i$, $I_{j\to i}(t)$, follows the equations of motion,

$$V_j(t) - V_i(t) = L\frac{I_{j\to i}(t+\Delta t) - I_{j\to i}(t)}{\Delta t} + R I_{j\to i}(t), \tag{S5}$$

$$\sum_{j\neq i}^{nn} I_{j\to i}(t) = C\frac{V_i(t+\Delta t) - V_i(t)}{\Delta t}. \tag{S6}$$

We used a measured waveform of the input THz electrical pulse (Fig. 1b of the main text) as the driving voltage at the injection site and simulated the time evolution of the plasmon electric field using Eqs. (S5) and (S6). Figure S9a shows a snapshot of the plasmon potential at a time delay of 2 ps. We used $C = 9.7 \times 10^{-4}$ Fm$^{-2}$ (same as in the velocity calculation) and $L = \hbar/4v_F e^2\sqrt{\pi n}$. The value of $R$ used in the simulation corresponds to the DC resistance measurement, as shown in Fig. S9b. Consequently, the waveform simulation requires no adjustable parameters. In addition to the initial pulse, we include the first echo pulse that arises from multiple reflections at the generation and detection PC switches. The echo pulse is calculated by multiplying the amplitude of the initial pulse by the reflection factors $r_g$ and $r_d$, and adding the time delays $\tau_g$ and $\tau_d$. Here, the subscripts $g$ and $d$ represent the generation and detection PC switches, respectively. We used $\tau_g$ = 6.14 ps and $\tau_d$ = 5.78 ps obtained from the known distance between the graphene and the two PC switches, along with the known $v_{CPW}$. For the reflection factors, we employed $r_g$ = $r_d$ = 0.2. The simulation results, plotted in Fig. 4a of the main text, show good agreement with the experimental waveform for the main and first echo pulse. Although accurately modeling the second and third echo pulses is expected to replicate the entire waveform, this is beyond the scope of our current study and purpose.

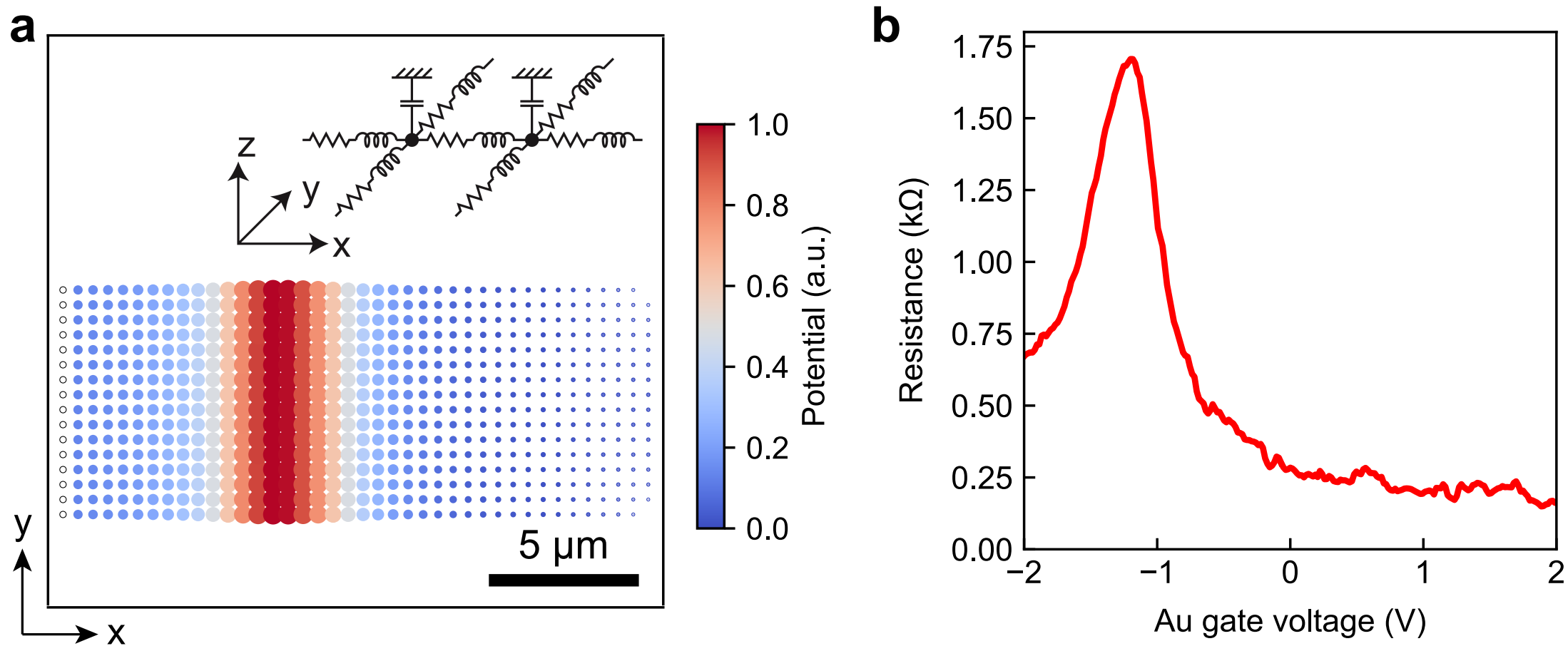

**Figure S9 | Waveform simulation of acoustic plasmon wavepackets. a**, Snapshot of plasmon transport at a carrier density of $n$ = 10.6 × $10^{11}$ $cm^{-2}$ and time delay of 2.0 ps. The size and color of each dot indicate the potential at individual nodes in the distributed constant circuit. The inset illustrates the elements, $R$, $L$, and $C$, constituting each node. The spacing between nodes is 0.5 μm. Plasmons are injected from the leftmost points, indicated by black open circles. Here, the echo pulse is excluded for clarity. **b**, DC resistance of graphene plotted as a function of the Au gate voltage.

The crossover between coherent plasmon transport and incoherent diffusive transport occurs at $k^2/LC \sim R^2/4L^2$. Eq. (S2) gives plasmon transport for $k^2/LC \gg R^2/4L^2$ and diffusive transport for $k^2/LC < R^2/4L^2$. In the diffusive transport regime, the peak of a solitary pulse remains at the injection sites. The temporal maximum of a diffusive wave at distance $x$ appears when $t = RC^2x^2/2$, and the velocity of this peak is slower than the plasmon velocity by a factor of $3/(R/2\sqrt{C/L}\ x)$.

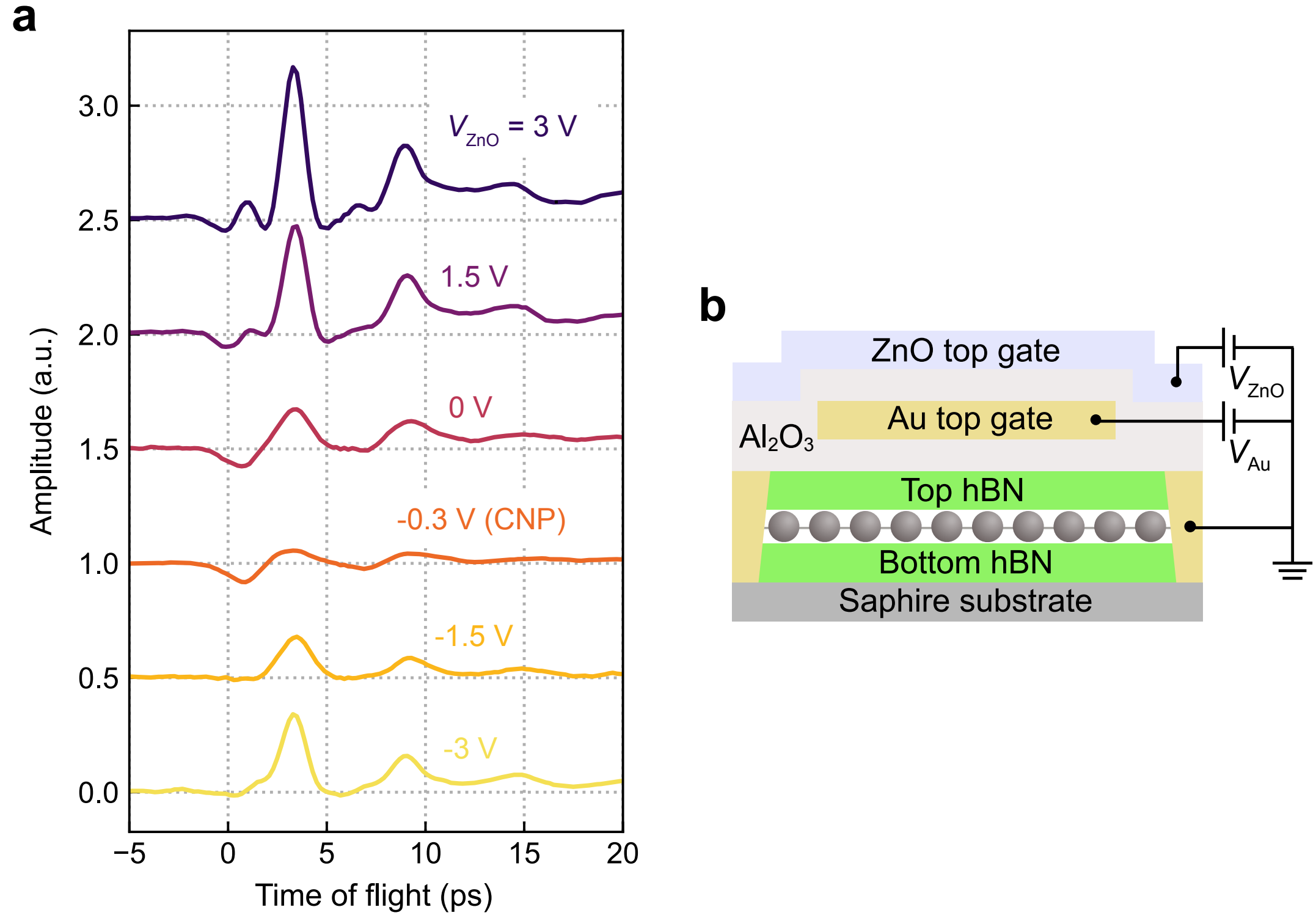

**Figure S10 | Coupling control to an acoustic plasmon. a,** Time domain waveforms of acoustic plasmons ($V_{Au}$ = 1.5 V) in the Au gate device for different values of $V_{ZnO}$. **b**, Cross-sectional view of the sample structure.

## V. Conductive and capacitive coupling to an acoustic plasmon

As shown in Fig. S10, the impedance matching between acoustic plasmons in graphene and the electrical pulse in the CPW can be controlled by tuning $V_{ZnO}$. Note that $V_{ZnO}$ does not affect the carrier density in graphene beneath the Au gate. At high carrier density ($|V_{ZnO}|$ = 3 V), the waveform takes on a unipolar shape. This indicates that the CPW and graphene are coupled conductively through the highly-doped ZnO gated area. Conversely, near the CNP ($V_{ZnO}$ = -0.3 V), the waveform becomes bipolar shape. This indicates that the coupling between the CPW and graphene becomes capacitive across the highly-resistive ZnO gated area. All measurements presented in the main text were performed at $V_{ZnO}$ = 3 V. It is important to note that the length of the ZnO gated area is sufficiently short to avoid any additional broadening of the input electrical pulse. This is confirmed by the observation of a 1.2 ps duration acoustic plasmon wavepacket (shown in Fig. 4a), which matches the pulse duration of the input electrical pulse (as seen in Fig. 1b).

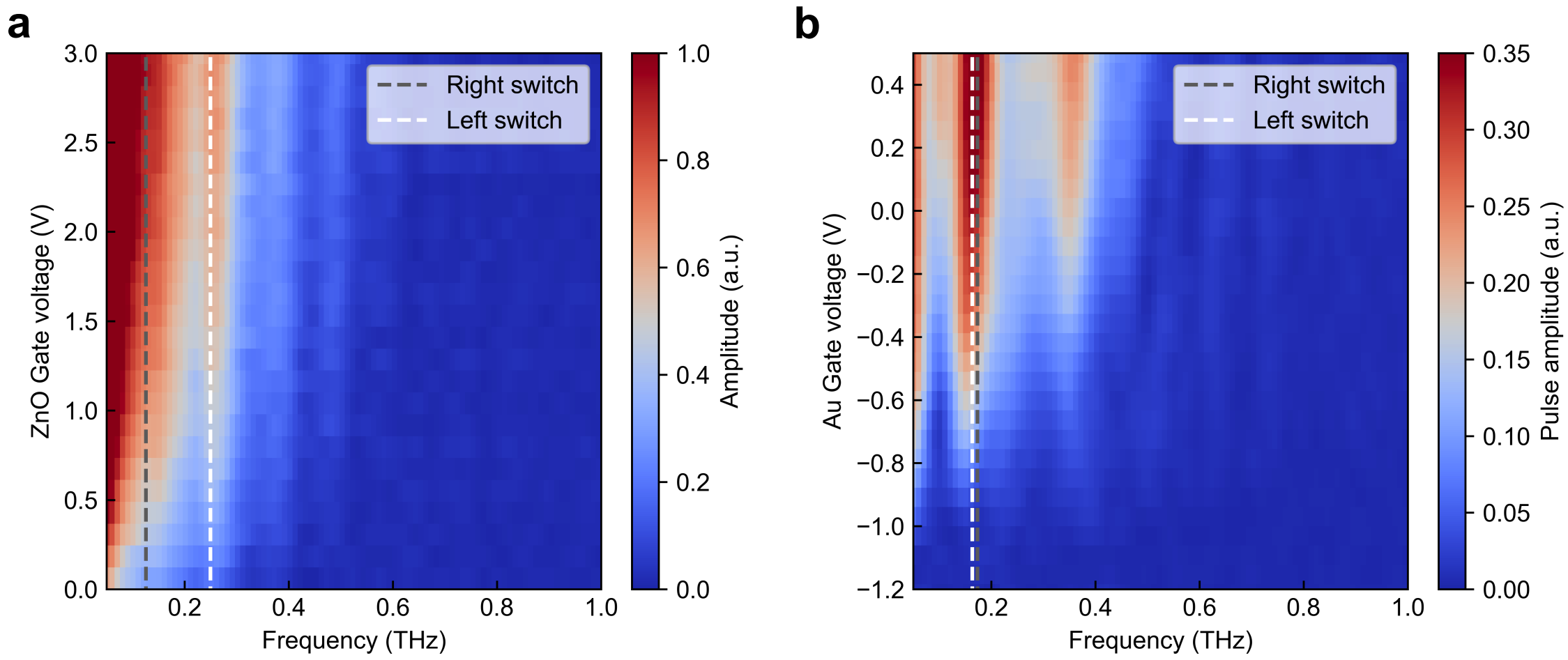

**Figure S11 | FFT spectra of graphene plasmon. a**, Unscreened plasmon. **b**, Acoustic plasmon. Expected resonance from the multiple reflection at left and right PC switches are shown by the dashed lines. These dashed lines are calculated based on the measured velocities and the circuit geometry.

## VI. Fourier Transform spectra of graphene plasmon wavepackets

Figure S11 presents the Fourier Transform (FFT) spectra for both unscreened and acoustic plasmons, plotted as a function of gate voltage. These spectra are derived from the data depicted in Figs. 2d and 3d of the main text. Given the broadband nature of our approach to generating graphene plasmons, not restricted to a fixed wavenumber by a cavity or antenna, the FFT spectra lack the typical features of plasmon resonance. Instead, they display a broadband plasmon component alongside multiple reflections from the THz circuit.

## VII. Transport properties of graphene FET

Figure S12 presents the resistance of graphene as a function of the gate voltage for both ZnO and Au gated devices. The carrier mobility of graphene ($\mu$) is calculated by fitting the experimental data to the equation $R = (ne\mu)^{-1} + R_b$[6], where $n$ represents carrier density, $e$ is the elementary charge, and $R_b$ accounts for background resistance including contact and interconnect resistances. Note that the mobility difference between the two devices arises from the variation of disorder level in graphene or subtle difference in the fabrication process and is irrelevant to the material of the gate electrodes.

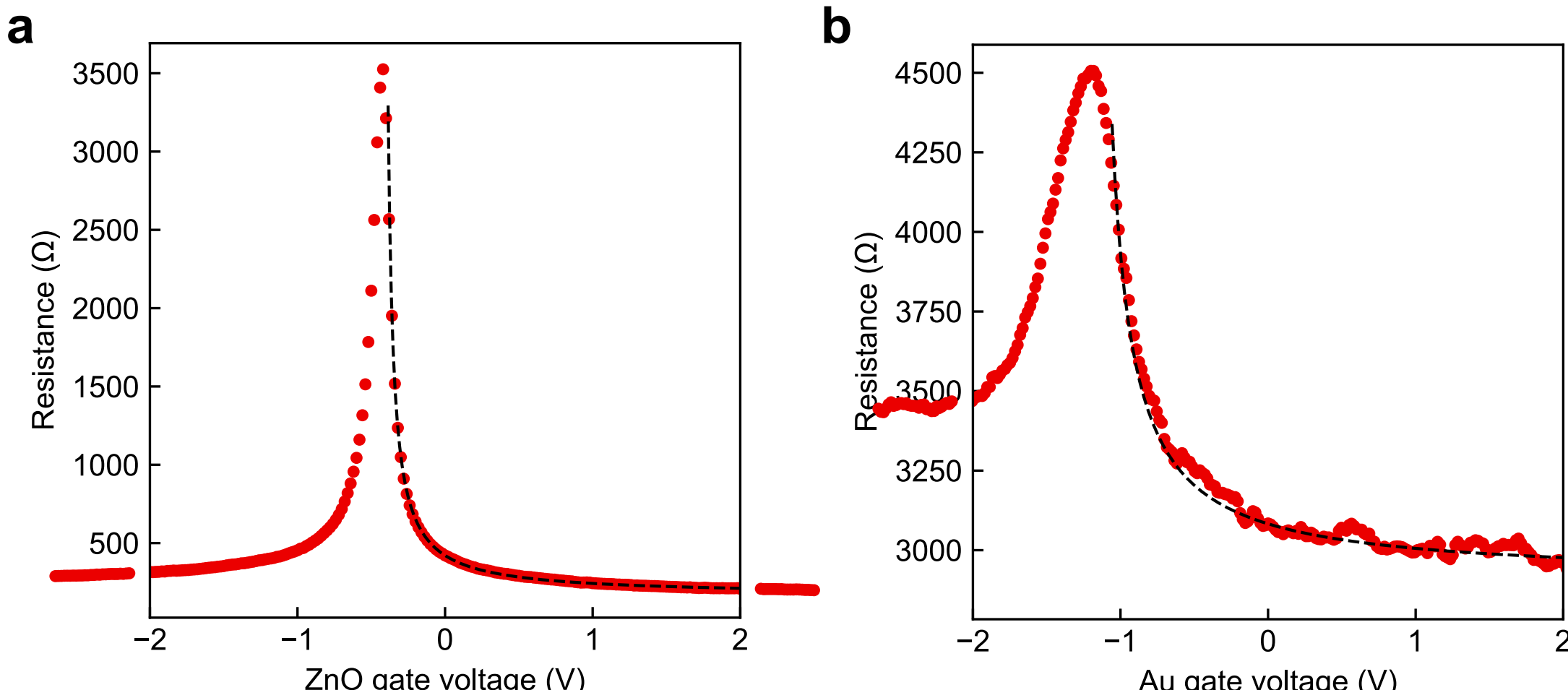

**Figure S12 | Gate voltage dependence of graphene resistance. a**, ZnO gated device. **b**, Au gated device. Black dashed curves show fitted results to determine the carrier mobility ($\mu$) of graphene with $\mu$ = 250,000 $cm^2V^{-1}s^{-1}$ for the ZnO gate device (**a**) and $\mu$ = 73,000 $cm^2V^{-1}s^{-1}$ for the Au gate device (**b**).